\documentclass[journal,comsoc]{IEEEtran}

\usepackage[T1]{fontenc}
\usepackage{aecompl} 
\usepackage{caption}

\ifCLASSINFOpdf
   \usepackage[pdftex]{graphicx}

   \graphicspath{{{\vartheta}_{k}^v/pdf/}{{\vartheta}_{k}^v/jpeg/}}

   \DeclareGraphicsExtensions{.pdf,.jpeg,.png}
\else

   \usepackage[dvips]{graphicx}

   \graphicspath{{{\vartheta}_{k}^v/eps/}}
  
   \DeclareGraphicsExtensions{-eps-converted-to.pdf}
\fi

\usepackage{amsmath}
\usepackage{subeqnarray}
\usepackage{cases}
\usepackage{cite}
\usepackage{graphicx}
\usepackage{subfigure}
\usepackage{tabularx,booktabs}
\usepackage{dingbat}
\usepackage{diagbox}
\usepackage[ruled,linesnumbered]{algorithm2e}
\usepackage{amssymb}
\usepackage{algorithmic}
\usepackage{esint}
\usepackage{color}
\usepackage{lipsum}
\usepackage{cuted}
\usepackage{stfloats}
\usepackage{ntheorem}
\usepackage{threeparttable}
\usepackage{bbm}
\usepackage{tabu}
\usepackage{multirow}
\usepackage{makecell}
\usepackage{colortbl}
\usepackage{url}
\usepackage{array}
\usepackage[colorlinks,
            linkcolor=black,       
            anchorcolor=black,  
            citecolor=black,
            urlcolor=black,        
            ]{hyperref}
\newcolumntype{C}{>{\centering\arraybackslash}X}
\theoremseparator{:}

\newtheorem{lemma}{Lemma}

\newtheorem*{Proof}{Proof}

\usepackage[cmintegrals]{newtxmath}

\begin{document}

\title{Age of Information in Energy Harvesting Aided Massive Multiple Access Networks}
\author{ Zhengru~Fang,~\IEEEmembership{ Student Member,~IEEE,}
Jingjing~Wang,~\IEEEmembership{Senior Member,~IEEE,}\\
Yong~Ren,~\IEEEmembership{Senior Member,~IEEE,}
Zhu~Han,~\IEEEmembership{Fellow,~IEEE,}\\
H.~Vincent~Poor,~\IEEEmembership{Life Fellow,~IEEE,}
and Lajos~Hanzo,~\IEEEmembership{Life Fellow,~IEEE}
}

\markboth{IEEE JOURNAL ON SELECTED AREAS IN COMMUNICATIONS}%
{Shell \MakeLowercase{\textit{et al.}}: Bare Demo of IEEEtran.cls for IEEE Communications Society Journals}

\maketitle
\newcommand\blfootnote[1]{%
\begingroup
\renewcommand\thefootnote{}\footnote{#1}%
\addtocounter{footnote}{-1}%
\endgroup
}
\blfootnote{This work was supported in part by National Key R\&D Program of China (No. 2020YFD0901000), in part by the National Natural Science Foundation of China (No. 62071268) and in part by the Young Elite Scientist Sponsorship Program by CAST (No. 2020QNRC001). The work of Z. Han was partially supported by NSF CNS-2128368, CNS-2107216, Toyota and Amazon. Moreover, L. Hanzo would like to acknowledge the financial support of the Engineering and Physical Sciences Research Council projects EP/P034284/1 and EP/P003990/1 (COALESCE) as well as of the European Research Council's Advanced Fellow Grant QuantCom (Grant No. 789028). \textit{(Corresponding author: Jingjing Wang.)}}
\blfootnote{Z. Fang and Y. Ren are with the Department of Electronic Engineering, Tsinghua University, Beijing, 100084, China. 
E-mail: fangzr19@mails.tsinghua.edu.cn, reny@tsinghua.edu.cn.}
\blfootnote{J. Wang is with the School of Cyber Science and Technology, Beihang University, Beijing 100191, China. Email: drwangjj@buaa.edu.cn.} 
\blfootnote{Z. Han is with the Department of Electrical and Computer Engineering in the University of Houston, Houston, TX 77004 USA, and also with the Department of Computer Science and Engineering, Kyung Hee University, Seoul, South Korea, 446-701. E-mail: hanzhu22@gmail.com.}
\blfootnote{H. V. Poor is with the Department of Electrical and Computer Engineering, Princeton University, Princeton, NJ 08544 USA. E-mail: poor@princeton.edu.}
\blfootnote{L. Hanzo is with the School of Electronics and Computer Science, the University of Southampton, Southampton SO17 1BJ, U.K. E-mail: lh@ecs.soton.ac.uk.}\par
\vspace{-0.3cm}
\begin{abstract}
  Given the proliferation of the massive machine type communication devices (MTCDs) in beyond 5G (B5G) wireless networks, energy harvesting (EH) aided next generation multiple access (NGMA) systems have drawn substantial attention in the context of energy-efficient data sensing and transmission. However, without adaptive time slot (TS) and power allocation schemes, NGMA systems relying on stochastic sampling instants might lead to tardy actions associated both with high age of information (AoI) as well as high power consumption. For mitigating the energy consumption, we exploit a pair of sleep-scheduling policies, namely the multiple vacation (MV) policy and start-up threshold (ST) policy, which are characterized in the context of three typical multiple access protocols, including time-division multiple access (TDMA), frequency-division multiple access (FDMA) and non-orthogonal multiple access (NOMA).
  Furthermore, we derive closed-form expressions for the MTCD system's peak AoI, which are formulated as the optimization objective under the constraints of EH power, status update rate and stability conditions. An exact linear search based algorithm is proposed for finding the optimal solution by fixing the status update rate. As a design alternative, a low complexity concave-convex procedure (CCP) is also formulated for finding a near-optimal solution relying on the original problem's transformation into a form represented by the difference of two convex problems.  
  Our simulation results show that the proposed algorithms are beneficial in terms of yielding a lower peak AoI at a low power consumption in the context of the multiple access protocols considered.
\end{abstract}
\IEEEpeerreviewmaketitle
\section{Introduction}
\subsection{Background}
\IEEEPARstart{M}{achine} type communication devices (MTCDs) are set to proliferate to 70+ billion in wireless networks by 2025\cite{chen2020massive}. As a further trend, compelling new applications, such as Industry 4.0, telemedicine and automatic driving are finding their way into the technical requirements of wireless systems\cite{wangthirty,yuan20205g}. Many of the above-mentioned applications are time-sensitive and/or power-constrained, especially the family of sophisticated Internet of Things (IoT) applications\cite{lu2015resource}. In this context next generation multiple access (NGMA) techniques play a critical role in terms of supporting large-scale user access as well as guaranteeing a high quality of experience.

For the sake of quantifying the `freshness' of the received data, the age of information (AoI) metric has been proposed and employed in a wide variety of wireless networks\cite{yates2021age}, which quantifies the time elapsed since the generation instant of the latest received status update at a destination\cite{abdel2019optimized}. In the IoT era, ubiquitous MTCDs are often deployed for monitoring dynamic environmental information\cite{liang2018cluster}. To elaborate, the data collected by MTCDs can be transmitted to so-called information fusion nodes in real-time for further analysis, control signal extraction as well as for remote data reconstruction, where the `freshness' of data directly influences the performance of real-time MTCD-aided wireless applications\cite{zhang2021pricing}. 

These MTCDs are often characterized by comparatively small amount data as well as by AoI-sensitive nature in typical data-collection oriented uplink (UL) communications. In order to maintain the freshness of the data, MTCDs have to periodically transmit their UL packets at a given frequency and data rate, which may result in the premature depletion of their limited batteries as well as in buffer-overflow\cite{tang2019energy}. From an economical and feasibility perspective, disposable batteries are inadequate for scalable long-term monitoring\cite{pei2020socially}. As a remedy, energy harvesting (EH) techniques constitute a low-cost renewable energy source for MTCDs, which gleans energy from the ambient radio frequency (RF) signals.

However, due to the inevitable signal attenuation and the stochastic nature of status updates, it remains a challenge to meet the AoI requirements in the context of time- and/or energy-sensitive applications. Specifically, in the open literature of massive access networks, most researchers primarily considered the network's throughput and power consumption. However, there is a paucity of literature on the AoI, which is a complex function of both throughput and power consumption. 
Hence it is imperative to design efficient multiple access protocols for AoI-sensitive EH-aided MTCD networks\cite{wu2018spectral}. For instance, EH-aided networks relying on time-division multiple access (TDMA)\cite{chi2019energy}, frequency-division multiple access (FDMA)\cite{mao2019energy} and non-orthogonal multiple access (NOMA)\cite{zhou2020computation} have been widely investigated both in terms of their power consumption and throughput, which are closely linked to their power control and resource allocation. In EH based multiple access networks, the customizable parameters typically include the transmission slot length, the status update rate and the EH parameters. In order to minimize the average of the system's peak AoI, we have to determine the optimal time-slot duration and power allocation under the specific power consumption constraint of the EH-aided MTCD networks considered.

\subsection{State-of-the-art}
In this subsection, we review the related literature of EH-based massive multiple access networks, with an emphasis on their energy-efficient resource allocation, massive multiple access protocols and AoI-oriented optimization schemes.\par
\textbf{(1) Energy-efficient resource allocation}: Numerous energy-efficient resource allocation schemes have been designed for EH\cite{hu2018lyapunov,xia2021online,singh2018toward,hu2018integrated,wang2019distributed}. To minimize the average outage probability at a finite battery capacity, Hu \textit{et al.} \cite{hu2018lyapunov} harnessed the classic Lyapunov optimization for transforming the long-term optimization problem into subproblems dedicated to each single TS. However, the drift-plus-penalty algorithm relying on Lyapunov optimization failed to obtain a desirable solution for devices equipped with small storage space. In \cite{xia2021online}, Xia \textit{et al.} formulated an EH-aided mobile edge computing based offloading system and solved an online distributed battery energy management optimization problem. However, their work did not consider the sleep-scheduling issues of small IoT devices. Considering an amplify-and-forward network relying on EH and relay nodes, Singh \textit{et al.} \cite{singh2018toward} creatively harnessed the relaying networks relying on EH source nodes and jointly optimized the power allocation and throughput by solving a non-convex problem. However, since the scenario only considered a single source node and a single relay node, the proposed methods may not apply to multiple access aided networks.\par
\textbf{(2) EH aided multiple access}: Qiao \textit{et al.} \cite{qiao2018achievable} optimized the throughput of an EH-based TDMA system with the aid of superposition coding in conjunction with fixed/variable decoding order. Vranas \textit{et al.} \cite{vranas2019gain} proposed a resource allocation scheme relying on a non-linear EH model for the uplink and they revealed that NOMA had a lower circuit power dissipation than TDMA, in the context of no more than 3 users. As a further development, Bouzinis \textit{et al.} \cite{bouzinis2020pareto} jointly optimized decentralized power transfer and radio access based on both NOMA and TDMA schemes. However, since the optimization of energy cost was ignored, the proposed system was unable to extend the lifetime of nodes. Moreover, Li \textit{et al.} \cite{li2021wireless} focused their attention on EH-aided mobile edge computing and solved the near-far problem of a NOMA-aided cooperative scheme, which may not be realistic, since it disregarded the switching cost of the time-division structure.\par
\textbf{(3) AoI-oriented optimization}: The AoI-oriented optimization of EH based networks has been widely investigated in the literature\cite{wu2017optimal,zheng2021age,krikidis2019average,azarhava2020age,bedewy2020optimizing,aydin2021energy,pan2020information,maatouk2019minimizing,chiariotti2021spectrum}. Specifically, Wu \textit{et al.} \cite{wu2017optimal} designed an optimal online status update scheme for minimizing the average AoI for various battery sizes in a continuous-time setting. To overcome the `selfishness' of the sensors and to formulate an attractive AoI upgrade strategy, Zheng \textit{et al.} \cite{zheng2021age} constructed a Stackelberg game based optimization problem and solved it via the classic Lagrangian method, where the worst case of the AoI was not considered. In \cite{krikidis2019average}, Krikidis \textit{et al.} studied a low-complexity EH regime, where a closed-form AoI expression was derived as a function of the storage capacitor's size. Additionally, Maatouk \textit{et al.}\cite{maatouk2019minimizing} investigated the AoI performance relying on NOMA and on orthogonal multiple access (OMA). Similarly, Chiariotti \textit{et al.} \cite{chiariotti2021spectrum} analyzed the AoI performance of TDMA and ALOHA-based schemes under both OMA and NOMA frameworks. However, the aforementioned contributions have not considered the fairness requirements of users.

\subsection{Motivation and New Contributions}
\begin{table*}[!t]
  \scriptsize
  \caption{Contrasting Our Contribution to the Litterature}
  \scriptsize
  \renewcommand{\arraystretch}{1}
  \centering
\begin{tabular}{|l|c|c|c|c|c|c|c|c|c|c|c|c|}
\hline
&\cite{tang2019energy}    & \cite{wu2018spectral} & \cite{qiao2018achievable}&\cite{bouzinis2020pareto} &\cite{li2021wireless}&\cite{zheng2021age} &\cite{krikidis2019average}&\cite{aydin2021energy} &\cite{azarhava2020age}&\cite{bedewy2020optimizing}& \cite{pan2020information} &Proposed work\\
\hline
\hline
Power allocation&\checkmark	&\checkmark	&\checkmark	&	&\checkmark	&\checkmark	&\checkmark	&\checkmark	&\checkmark	&&		&\checkmark\\
\hline
TS allocation&\checkmark	&\checkmark	&	&\checkmark	&\checkmark	&\checkmark	&	&	&\checkmark	&\checkmark&		&\checkmark\\
\hline
TDMA protocol &		&\checkmark	&\checkmark	&\checkmark	&	&	&	&	&\checkmark	&	&\checkmark&\checkmark\\
\hline
FDMA protocol &	&	&	&	&	&	&	&	&	&	&\checkmark&\checkmark\\
\hline
NOMA protocol&\checkmark	&\checkmark	&	&\checkmark	&\checkmark	&	&	&	&\checkmark	&	&	&\checkmark\\
\hline
AoI optimization & 		&	&	&	&	&\checkmark	&\checkmark	&\checkmark	&\checkmark	&\checkmark	&\checkmark&\checkmark\\
\hline
Sleep-scheduling& &	&	&	&	&	&	&\checkmark	&	&\checkmark	&	&\checkmark	\\
\hline
\end{tabular}\label{table_literature}
\end{table*}
Most of the aforementioned teatises assume that the transmitters of the EH aided networks are able to operate sustainably without any sleep-scheduling policy. By contrast, in many realistic scenarios, MTCDs have to rely on sleep-scheduling to avoid excessive power dissipation. However, the additional queuing delay caused by the sleep-scheduling policy might degrade the AoI performance of our system. Furthermore, we assume that the status updates are random. For instance, the randomness of data generation may arise from the constraints of energy harvesting systems relying on a limited battery, or due to an application-specific feature that needs random data sampling, hence affecting both the peak AoI and the overall power consumption. In contrast to the previous studies only focusing on throughput optimization, in this contribution, we propose an AoI-optimal and energy-efficient resource allocation scheme for EH-aided massive multiple access networks, beneficially optimizing the TS association of data transmission, energy harvesting and the system's idle mode. Against this above backdrop, our main contributions are as follows:
\begin{itemize}
  \item To the best of our knowledge, this is the first contribution optimizing both the peak AoI and the power consumption of EH aided large-scale multiple access networks, where the MTCDs adaptively switch between their active and idle mode in different time slots (TS). Under the assumption of stochastic status updates, we utilize a pair of different sleep-scheduling policies, namely the multiple vacation (MV) policy and the start-up threshold (ST) policy defined in\cite{tian2006vacation} for the sake of mitigating the power consumption, while relying on a queueing system. Then, we derive a closed-form expression for the average of the users' peak AoI in three different access schemes. 
  \item We design a beneficial TS allocation method for striking a trade-off between the average of the users' peak AoI and the power consumption. To solve the original non-convex optimization problem, we decompose it into a set of convex subproblems. By relying on an exact linear search method, the optimal energy and TS allocation scheme is obtained. Moreover, by converting the original problem into a typical difference of convex (DC) programming problem, we find its near-optimal solution by the convex-concave procedure (CCP) at a low complexity.
  \item Numerical results show that both the MV policy and the ST policy are capable of reducing the power consumption compared to the benchmark schemes. As for different access schemes, we may conclude that the MV-based NOMA policy outperforms all the others in terms of both the users' average peak AoI and power consumption. 
\end{itemize}
Our new contributions are boldly and explicitly contrasted to the literature at a glance in \mbox{Table \ref{table_literature}}.
\subsection{Organization}
The remainder of this article is structured as follows. The system model is detailed in Section \ref{sec:System_model}. In Section \ref{sec:Peak AoI Analysis in Multiple Access Networks}, closed-form expressions for the average peak AoI based on different multiple access protocols are derived. The associated non-convex problems and their solutions are investigated in Section \ref{Problem Formulation}. In Section \ref{sec:Simulation Results And Discussions}, we provided the performance analysis of the proposed schemes relying on different sleep-scheduling policies, followed by our conclusions in Section \ref{sec:Conclusion}.

\section{System Model}
\label{sec:System_model}

\begin{table}
  \caption{{\color{black}{Summary of Key Notations} }}
  \label{table:notation}
  \begin{center}
  \renewcommand{\arraystretch}{1.3}
  \begin{tabular}{c  p{6cm} }
  \hline
  {\color{black}{{\bf Notation}}} & {\hspace{2.5cm}}{\bf {\color{black}{Definition}}}
  \\
  \hline
  \textcolor{black}{$N$, $M$} & \textcolor{black}{Total number of MTCDs, start-up threshold for the ST policy} \\
  \textcolor{black}{$\tau_{p}$, $\tau_{t}$} & \textcolor{black}{Energy harvesting slot, transmission slot} \\
  \textcolor{black}{$\tau_{s}$, $\tau_{sc}$} & \textcolor{black}{Duration of the idle mode and of the switching cost}\\
  \textcolor{black}{$h_{p,n}$, $h_{t,n}$} & \textcolor{black}{Channel coefficient between the power station and the MTCD, channel coefficient between the BS and the MTCD} \\
  \textcolor{black}{$\varPhi _{t,n}$, $ \varPhi _{th}$} &  \textcolor{black}{Transmit power of the MTCD, upper bound of the power station’s charging power} \\
  \textcolor{black}{$\overline{\varPhi}$, $\varPhi_{w}$} & \textcolor{black}{Average power consumption of the MTCD, power consumption in the active mode} \\
  \textcolor{black}{$\varPhi_{s}$, $\varPhi_{sc}$} & \textcolor{black}{Power consumption in the idle mode, power consumption of the switching cost}\\
  \textcolor{black}{$B$, $N_0$}& \textcolor{black}{Available bandwidth, noise spectral density}\\
  \textcolor{black}{$R_{ub,n}$, $C_n$} & \textcolor{black}{Upper bound of channel capacity, actual average throughput}\\
  \textcolor{black}{${d}_n^{M-B}$, $L$}& \textcolor{black}{Distance of a pair of the BS-MTCD link, length of each packet} \\
  \textcolor{black}{$G_n$, $D_n $} & \textcolor{black}{Interval of the data generation, system delay of each packet} \\
  \textcolor{black}{$\lambda_n$, $\varUpsilon _{n}$} & \textcolor{black}{Packet arrival rate, energy efficiency}\\
  \textcolor{black}{$A_p^{MV}(\cdot)$, $A_p^{ST}(\cdot)$} & \textcolor{black}{Average peak AoI function for the MV policy and for the ST policy}\\
  \textcolor{black}{$\overline{A}$, $A_{p}$} & \textcolor{black}{Average AoI, peak AoI} \\
  \hline
  \end{tabular}
  \end{center}
  \end{table}%
    
  \begin{table}
  \caption{\textcolor{black}{Summary of Key Acronyms}} 
  \label{table:acronyms}
  \begin{center}
  \renewcommand{\arraystretch}{1.3}
  \begin{tabular}{c  p{6cm} }
  \hline
  {\bf \textcolor{black}{Acronym}} & {\hspace{2.5cm}}{\bf \textcolor{black}{Definition}}
  \\
  \hline
  \textcolor{black}{MV, ST} & \textcolor{black}{Multiple vacation policy, start-up threshold policy} \\
  \textcolor{black}{TM, TS} & \textcolor{black}{Time-division multiple access (TDMA) protocol with the MV policy, TDMA protocol with the ST policy} \\
  \textcolor{black}{FM, FS} & \textcolor{black}{Frequency-division multiple access (FDMA) protocol with the MV policy, FDMA protocol with the ST policy} \\
  \textcolor{black}{$NM$, $NS$} & \textcolor{black}{Power-domain non-orthogonal multiple access (PD-NOMA) protocol with the MV policy, PD-NOMA protocol with the ST policy}\\ 
  \hline
  \end{tabular}
  \end{center}\vspace{-0.63cm}
  \end{table}%

\subsection{Network Architecture}
\label{sec:Network Architecture}
\begin{figure}[t]
  \centering
  \includegraphics[width=0.47\textwidth]{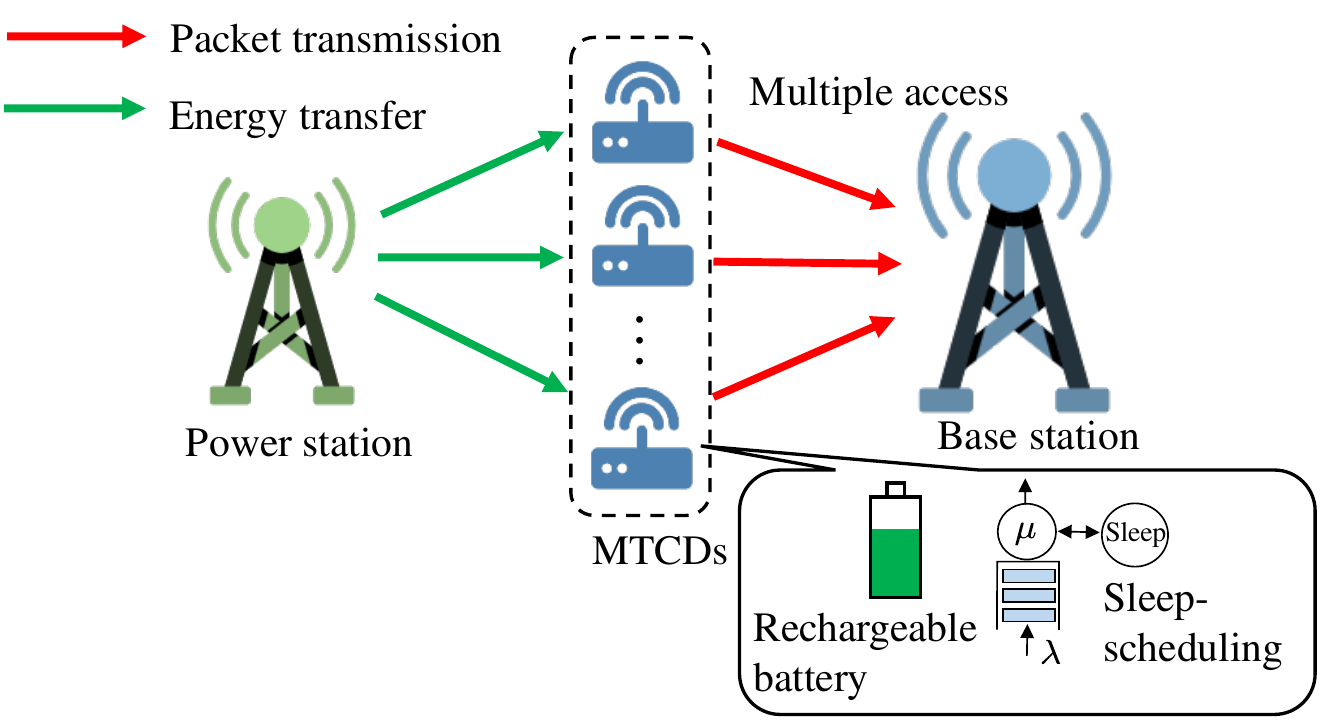}
  \caption{EH-aided massive access MTCD network.}
  \label{fig:model}
  \vspace{-6mm}
\end{figure}

The EH-aided massive multiple access network considered is portrayed in Fig. \ref{fig:model}, which consists of a wireless power station as an EH source, a base station (BS) for data collection and $N$ MTCDs denoted by $n\in\mathcal{N}\stackrel{\triangle}{=}\{1,2,\cdots,N\}$. Equipped with a single antenna, each MTCD senses the surrounding environment for collecting information and transmits it to the BS. Moreover, the power station recharges the MTCDs, which are also equipped with RF-based EH modules. Bearing in mind the independence of each data packet, we assume that all the status updates of each MTCD obeys a Poissonian process. For the sake of reducing the power consumption, each MTCDs relies on a specific sleeping scheduling, i.e. each MTCD turns the transmit power off, if its data transmission queue is empty\footnote{When the transmitter is active or in idle mode, the latest data is queued instead of being transferred. When the queue is empty, the MTCD enters into idle mode.}. To reduce the co-channel interference, the procedure of EH and data transmission of each MTCD is operated in separate slots. Thus, each MTCD has an \textbf{active mode} and \textbf{idle mode}\footnote{For each mode, the environment sensing is uninterrupted, relying on MTCDs' low-power dissipation sensors.}, Furthermore, the active mode is partitioned into two slots, namely the EH slot $\tau_{p}$ and the transmission slot $\tau_t$. In this paper, TDMA, FDMA and NOMA are investigated in our EH multiple access system. For readability, the important notations and acronyms used in this paper are listed in Table \ref{table:notation} and Table \ref{table:acronyms}, respectively.

\subsection{Energy Harvesting Model}
\label{sec:Energy Harvesting Model}
In our model, a power station charges the MTCDs relying on EH techniques. Let $\tau_{p,n}^{\left[ i \right]}$ denote the duration of EH between the power station and the $n$th MTCD, where $\left[ i \right]$ represents the index of the status update. Moreover, let the constant $h_{p,n}\left(t\right)$ be the channel coefficient of the wireless links between the power station and the $n$th MTCD. At the time instant $t$, the EH power of the $n$th MTCD's receiver can be formulated as
\begin{equation}\label{eq:EH_power}
  \begin{aligned}
    \varPhi _{r,n}^{\left[ i \right]}\left(t\right)=\min \left\{ \eta _k\varPhi _{wp,n}^{\left[ i \right]} h_{p,n} ,\varPhi _{th} \right\}, \ \forall n \in \mathcal{N}, 
  \end{aligned}
\end{equation}
where $0<\eta_p <1$ is the EH efficiency, which substantially depends on the specific hardware components. Furthermore, $\varPhi _{wp,n}^{\left[ i \right]}\left(t\right)$ denotes the charging power of the power station directed towards the $n$th MTCD, while $\varPhi _{th}$ is the upper bound of the power station's charging power. Let $\tau_p^{[i]}$ be the duration of energy harvesting for the $i$th status update. Thus, the energy received by the MTCD from the power station for a single status update is formulated as:
\begin{equation}\label{eq:EH_energy}
  \begin{aligned}
    E_{h}^{[i]}=\int_0^{\tau _{p}^{[i]}}{\varPhi _{r,n}^{\left[ i \right]}\left(t\right)dt}, \ \forall n \in \mathcal{N}.
  \end{aligned}
\end{equation}
Furthermore, the upper-bound of the battery capacity in the MTCD is set to $C_{ub}$ which has the unit of Joule (J). While the MTCD is activated for the $i$th status update, the conservation of energy requires that the energy received by the MTCD must satisfy $E_{h}^{[i]}=C_{ub}$. 
\subsection{Transmission Model and Channel Analysis}
\label{sec:Transmission Model}
The most energy harvested by the MTCD from the BS is used for data transmission. Let $\varPhi _{t,n}^{\left[ i \right]}$ and $\varPhi _{a,n}^{\left[ i \right]}$ denote the power of data transmission, and that of active state in the $i$th status update, respectively. Furthermore, the associated battery cost values are $C_t$ and $C_a$, respectively. Therefore, the energy constraint of the MTCD for the $i$th status update yields $\int_0^{\tau _{t}^{[i]}}{\varPhi _{t,n}^{\left[ i \right]}\left(t\right)dt}=C_t$ and $\int_0^{\tau _{t}^{[i]}}{\varPhi _{a,n}^{\left[ i \right]}\left(t\right)dt}=C_a$. 
Likewise, the energy constraint of the MTCD for the idle mode is given by $\int_0^{\tau _{s}^{[i]}}{\varPhi _{s,n}^{\left[ i \right]}\left(t\right)dt}=C_s$. Moreover, we assume that $C_{ub}=C_t+C_a+\varepsilon C_s$, where $\varepsilon>0$ denotes the coefficient of EH for the idle mode. In the scenario considered, the MTCD only transmits a single packet having the length of $L$ bits in a transmission time slot, and we assume that all MTCDs have the same transmission rate. Hence, the upper bound of channel capacity should satisfy:
\begin{equation}\label{eq:capacity cons}
  \begin{aligned}
  R_{ub,n}=\frac{\left(\mathcal{J}+1\right) L}{\tau _{t}^{[i]}}>\mathcal{C} _n,
\end{aligned}
\end{equation}
where $\mathcal{C} _n=\frac{L}{\tau _{b}^{[i]}}$ is the actual average throughput of the $n$th MTCD, and $\mathcal{J}>0$ denotes the gap between the ideal capacity and practical throughput. In a Gaussian multiple-access channel, $N$ independent MTCDs simultaneously transmit their status updates to the BS relying on the assumption of the additive white Gaussian noise (AWGN).\par
\textbf{(1) TDMA:} When relying on TDMA, we can allocate the complete bandwidth $B$ of the channel to all MTCDs, but these MTCDs only transmit packets for a limited time period. Therefore, the upper bound of the channel capacity relying on TDMA for the $n$th MTCD is given by:
\begin{equation}\label{eq:capacity TDMA}
  \begin{aligned}
    R_{ub,n}^{T,\left[ i \right]}\left( t \right) =\frac{B\tau _t^{[i]}}{\tau _b^{[i]}}\log _2\left( 1+\frac{\varPhi _{t,n}^{\left[ i \right]}\left(t\right)\left| 
      h_{t,n}\left( t \right) \right|^2}{BN_0} \right),
\end{aligned}
\end{equation}
where $\left| 
h_{t,n}\left( t \right) \right|=10^{-3}\theta_0 d_n^{-a_0}$ is the channel coefficient of the wireless link between the BS and the $n$th MTCD at time $t$, while $\theta_0$ and $d_n$ denote the small-scale fading parameter and the link length, respectively. Let $a_0$ denote the path loss of the link, $B$ represents the available bandwidth and $N_0$ is the noise spectral density. \\
\textbf{(2) FDMA:} As for FDMA, the available bandwidth is partitioned into $N$ subchannels, and the associated channel capacity of the $n$th MTCD is given by:
\begin{equation}\label{eq:capacity FDMA}
  \begin{aligned}
    R_{ub,n}^{F,\left[ i \right]}\left( t \right) =\frac{B\tau _t^{[i]}}{N\tau _b^{[i]}}\log _2\left( 1+\frac{\varPhi _{t,n}^{\left[ i \right]}\left(t\right)\left| 
      h_{t,n}\left( t \right) \right|^2}{(B/N)N_0} \right).
\end{aligned}
\end{equation}
\textbf{(3) NOMA:} As for power-domain NOMA (PD-NOMA), we assume that $h_{t,1}\left(t\right) \geq h_{t,2}\left(t\right) \geq\cdots\geq h_{t,N}\left(t\right) $. In order to ensure the successful detection of the NOMA signal, we should control the transmit power of each MTCD to satisfy the following conditions for successive interference cancellation (SIC) at the BS's receiver:
\begin{equation}\label{eq:noma-2}
  \begin{aligned}
  {\varPhi _{t,j}^{\left[ i \right]}\left(t\right)\gamma_j}-\sum_{k=j+1}^N{{\varPhi _{t,k}^{\left[ i \right]}\left(t\right)\gamma_k\left(t\right)}}\geqslant \varPhi_{th}  \ \ j=1,2,\cdots,N-1,
\end{aligned}
\end{equation}
where $\gamma_n\left(t\right)=\frac{\left| 
h_{t,n}\left( t \right) \right|^2}{BN_0}$, while $\varPhi_{th}$ is the minimum power required for successfully decoding the NOMA signal received. Therefore, the signal to interference plus noise ratio (SINR) for the $n$th MTCD can be formulated as follows:
\begin{equation}\label{eq:noma-3}
  \begin{aligned}
    \varTheta _n^{\left[ i \right]}\left(t\right)=\begin{cases}
      \frac{\varPhi _{t,n}^{\left[ i \right]}\left(t\right)\gamma _n\left(t\right)}{1+\sum_{j=n+1}^N{\varPhi _{t,j}^{\left[ i \right]}\left(t\right)\gamma _j\left(t\right)}} & 1\leq n \leq N-1,\\
      \varPhi _{t,n}^{\left[ i \right]}\left(t\right)\gamma _n\left(t\right) & n=N.\\
    \end{cases}
\end{aligned}
\end{equation}
Because only $\left(\tau_{t}^{[i]}/\tau_{b}^{[i]}\right)\times 100 \%$ TSs are regarded as the legitimate transmission duration, the average transmission rate of the $n$th MTCD is:
\begin{equation}\label{eq:noma-4}
  \begin{aligned}
R_{ub,n}^{N,[i]}=\frac{B\tau _t^{[i]}}{\tau _b^{[i]}}\log _2\left( 1+\varTheta _n^{[i]} \right).
\end{aligned}
\end{equation}
Note that the channel state information (CSI) is assumed to be perfectly known both at the BS and at all the MTCDs. Then the CSI of BS can be estimated relying on the downlink pilots at the user equipment (UE), while the CSI of the MTCDs can be acquired through uplink channel estimation by the BS. Furthermore, since the packet length $L$ is typically small and the time between status updates is short, the channel can be regarded as having a near-constant slow- and fast-fading\cite{zheng2021age}.
\section{Peak AoI Model}
\label{sec:Peak AoI Analysis in Multiple Access Networks}
\begin{figure}[t]
  \centering
  \includegraphics[width=0.46\textwidth]{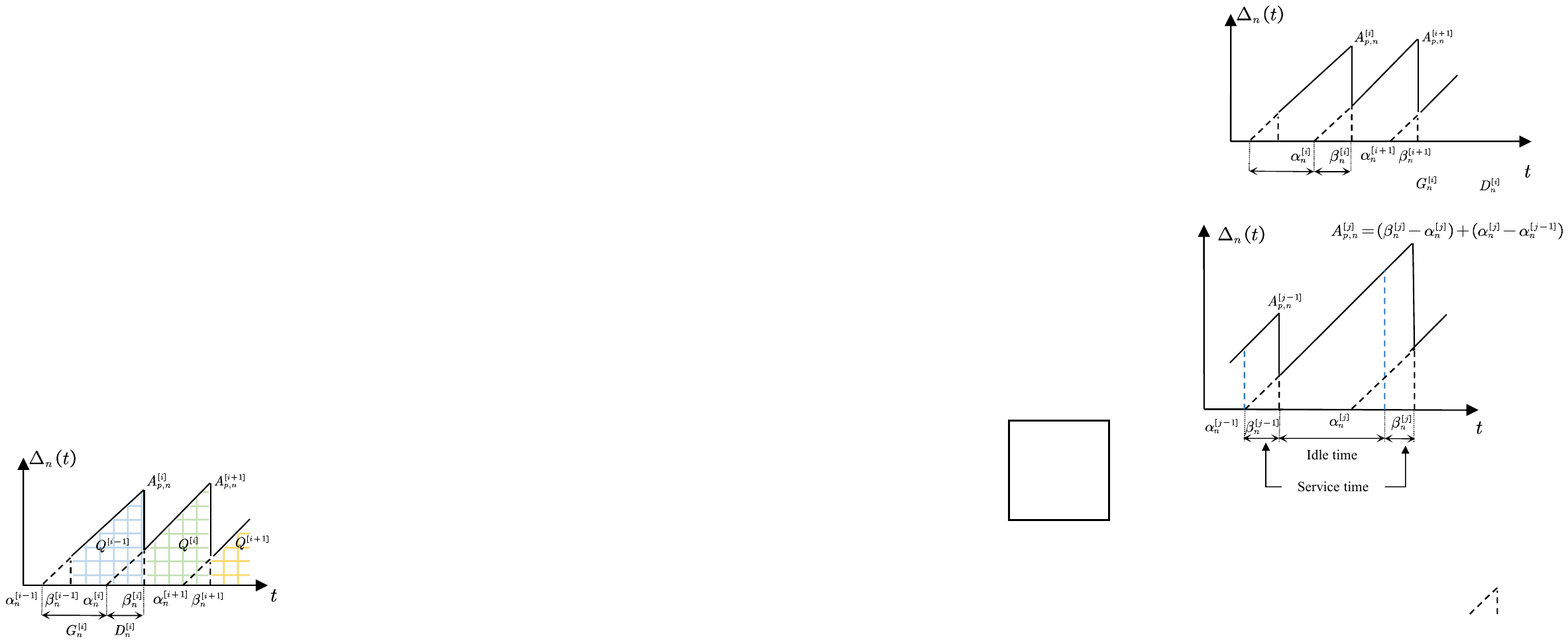}
  \caption{Age of information of first-in-first-out (FIFO) single queue.}
  \label{fig:paoi}
  \vspace{-6mm}
\end{figure}
In this section, we carry out a closed-form worst case analysis of the system's AoI based on queueing theory in terms of three multiple access protocols, namely TDMA, FDMA and NOMA. Additionally, we use different vacation queues relying on a single server to model a pair of sleep-scheduling policies. {\color{black}{The AoI of the $n$th MTCD of the $i$th status update is illustrated in Fig. \ref{fig:paoi}, which can be defined by
\begin{equation}\label{eq:AoI}
  \begin{aligned}
    {A} _n^{[i]}\left( t \right) =t-\alpha^{[i]} _n\left( t \right),
  \end{aligned}
\end{equation}
where $\alpha _n^{[i]}\left( t \right)=\alpha^{[i]} _n$ represents the generation moment of the latest status information before $t$ in the BS, while $\alpha_n^{\left[ i \right]}$ and $\beta_n^{\left[ i \right]}$ denote the generation and transmission moment of the $i$th status update, respectively\footnote{
  Taking into account the propagation speed of RF signal, the latency of the transmission between the transmitter and BS can be negligible in comparison to the system delay of the queue in each MTCD.}.}} Thus, we assume $\alpha_n^{[i]}<\beta_n^{[i]}$, $\alpha_n^{\left[ i \right]}\leq\alpha_n^{[i+1]}$ and $\beta_n^{\left[ i \right]}\leq\beta_n^{[i+1]}$. Moreover, $ A_{p,n}^{\left[ i \right]} $ is the instantaneous peak AoI of the $n$th MTCD. In the following section, the superscript ``$[i]$'' of the notations and $\left(t\right)$ are omitted for simplicity, because the optimization procedure for each status update is identical. Hence, $i$ can be omitted in the following equation:
\begin{equation}\label{eq:PAoI}
  \begin{aligned}
    \mathbb{E} \left[ A_p \right] &=\mathbb{E} \left[ \beta_n^{\left[ i \right]}-\alpha_n^{\left[ i-1 \right]} \right]=\mathbb{E} \left[ G_{n} \right] +\mathbb{E} \left[ D_{n} \right].
  \end{aligned}
\end{equation}
The average interval of the data generation can be expressed as $\mathbb{E} \left[ G_n \right] =\lim_{T\rightarrow\infty}T/\left(\sum_{i=1}^{\infty}\mathbbm{1}_{\{\alpha_n^{\left[ i \right]}<T\}}\right)$, where $\mathbbm{1}_{\{\cdot \}}$ is an indicator function, while $\mathbb{E} \left[D_{n}\right]$ denotes the average system delay of each packet\footnote{{\color{black}{The system delay contains the sum of waiting time, transmission time and idle period, while ``average'' indicates that we only consider their time-averaged value under the queue's stable condition.}}}. {\color{black}{Additionally, we use ``per-packet AoI'' to evaluate the performing of AoI associated with per packet in EH aided wireless networks. According to Section II-C in \cite{zheng2021age}, the ``per-packet AoI'' of the $i$th status update can be obtained by calculating the area of $Q^{[i]}$ in Fig. \ref{fig:paoi} as follows:
\begin{equation}\label{eq:per-packet AoI}
  \begin{aligned}
\overline{{A}}_n^{\left[ i \right]}&=\frac{Q^{\left[ i \right]}}{\beta _{n}^{\left[ i \right]}-\beta _{n}^{\left[ i-1 \right]}}
=\frac{Q^{\left[ i \right]}}{2\left( G_{n}^{\left[ i+1 \right]}+D_{n}^{\left[ i+1 \right]}-D_{n}^{\left[ i \right]} \right)}
=\frac{ D_n  + A_p }{2}.
  \end{aligned}
\end{equation}
Moreover, the expected value of the per-packet AoI is given by $\mathbb{E} \left[ \overline{A}_n \right]=\frac{\mathbb{E} \left[ D_n \right] +\mathbb{E} \left[ A_p \right]}{2}$, which reflects the time average AoI performance.}} In the following section, we further analyze the queueing system with heavily sleep-scheduling policies in a first-in-first-out (FIFO) manner\footnote{{\color{black}{Although last-in first-out (LIFO) queues are well-known to minimize AoI in many situations, FIFO queues are the standard queueing discipline in most wireless communication systems, especially when the devices are used to achieve multiple services and not only status updates.}}}.
\subsection{TDMA Protocol}
\label{sec:PAoI_TDMA-IB}
\begin{figure}[t]
  \centering
  \includegraphics[width=0.47\textwidth]{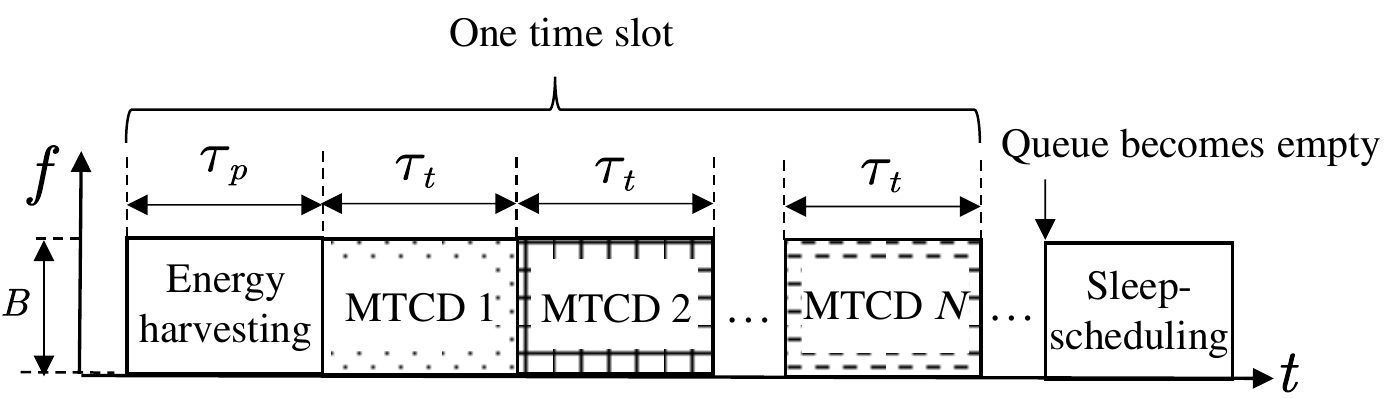}
  \caption{An example for illustrating the TDMA with sleep-scheduling.}
  \label{fig:TDMA-1}
  \vspace{-6mm}
\end{figure}
\begin{table*}[t]
  \arrayrulecolor{black}
  \caption{{\color{black}{The Definition of Different Sleep-Scheduling Policies.}}}
  \centering
  \label{policy_table}
  \begin{tabular}{|c|l|l|}
  \hline
  
  \textbf{{\color{black}{Sleep-scheduling policy}}} & \multicolumn{1}{c|}{\textbf{{\color{black}{Difference}}}}                                                                               & \multicolumn{1}{c|}{\textbf{{\color{black}{Similarity}}}}                                                                            \\ \hline\hline
  {\color{black}{MV policy}}               & \makecell[l]{{\color{black}{MTCD keeps on idle mode until it find at least}} \\{\color{black}{one packet waiting in the queue at the end of}} \\{\color{black}{one idle period.}}} & \multirow{2}{*}{\makecell[l]{{\color{black}{If the queue becomes empty, the}} \\{\color{black}{MTCD turns off its transmitter}} \\{\color{black}{and switches to idle mode.}}}} \\ \cline{1-2}
  {\color{black}{ST policy }}              & \makecell[l]{{\color{black}{MTCD keeps on idle mode until it find at least}} \\$M$ {\color{black}{packets waiting in the queue.}}}                                &                                                                                                            \\ \hline
  \end{tabular}
  \end{table*}

\begin{figure}[t]
  \centering
  \includegraphics[width=0.27\textwidth]{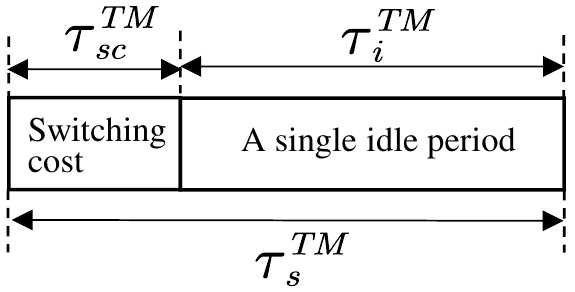}
  \caption{Illustrative example of sleep-scheduling slot with switching cost in MV policy.}
  \label{fig:Switch_time_slot}
  \vspace{-6mm}
\end{figure}
In TDMA, several MTCDs share the same carrier frequency channel by partitioning the transmission frame into multiple orthogonal TSs. In terms of the energy consumption of battery-limited MTCDs, it is unwise to maintain the power-on state of the device. Hence, we conceive a TDMA protocol based on sleep-scheduling, as shown in \mbox{Fig. \ref{fig:TDMA-1}.} In TDMA, all MTCDs share the entire bandwidth $B$ for transmitting their packets. It can be observed that a TDMA time slot is divided into $(N+1)$ TSs, which contain a single EH TS and $N$ transmission TSs. Specifically, all MTCDs harvest energy in the first TS of duration $\tau_p$. Moreover, each MTCD occupies a single TS of duration $\tau_{t}$. When the queue in the MTCD is empty, the transmitter is turned off to conserve energy. As shown in Table \ref{policy_table}, the main difference between MV policy and ST policy is the condition of queue starting transmission. To differentiate these two policies, we use the superscripts $TM$ and $TS$ to denote the parameters of TDMA associated with the MV and ST policy, respectively.\par
\textbf{(1) TDMA Using the MV Policy}: {\color{black}{To further analyze the corresponding model, we first introduce the concept of MV policy. \textbf{Specifically, if the MTCD is woken up from its idle mode and finds that the queue remains empty, it returns to its idle mode for the duration of $\tau_s^{TM}$, which can be calculated by the sum of the switching cost $\tau_{sc}^{TM}$ and the idle period $\tau_i^{TM}$. Furthermore, the illustration of the sleep-scheduling slot along with the switching cost of our MV policy is shown in Fig. \ref{fig:Switch_time_slot}
\footnote{In MV policy, MTCD needs a switching cost $\tau_{sc}$ before MTCD starts an idle period $\tau_{i}$. Switching cost is consumed for the monitoring of the status updates in the MTCD's queue.}. If however there is at least one packet in the queue during the switching cost slot, the MTCD starts to transmit its status updates.}}} We assume that the generation of the status update of the $n$th MTCD follows a Poissonian process with arrival rate $\lambda_n$, and its transmission time is $\tau_{t}^{TM}$, while the duration of each TDMA time slot without idle period is $\tau_b^{TM}=\tau_p +N\tau_{t}^{TM}$. Therefore, the procedure of status update of the $n$th MTCD using the TDMA protocol can be modeled by a MV aided queue defined in\cite{tian2006vacation}
\footnote{To streamline the terminology, we refer to the vacation period as an idle period.}. 
\begin{lemma}\label{lemma:MD1-mv}
  {\color{black}{If the arrival interval yields an exponential distribution and its service time is a constant, the queueing system can be regarded as a classical M/D/1 queue. Therefore, it may be readily shown that the p.g.f. of the number of packets when the packet depart in a FIFO manner as follows:}}
\begin{equation}\label{eq:pgf_L}
  \begin{aligned}
    \varPi \left( z \right)=\frac{\left( 1-\lambda\tau  \right) \left( 1-z \right) e^{-\left( \lambda-\lambda z \right) \tau }}{e^{-\left( \lambda-\lambda z \right) \tau }-z}.
\end{aligned}
\end{equation}
The average queue length may be expressed as
\begin{equation}\label{eq:L}
  \begin{aligned}
    \mathbb{E} \left[ \varPi \right]=\lambda \tau+\frac{\left( \lambda\tau  \right) ^2}{2\left( 1-\lambda\tau  \right)},
\end{aligned}
\end{equation}
while the average system delay may be formulated as follows:
\begin{equation}\label{eq:D0}
  \begin{aligned}
    \mathbb{E} \left[ D_0 \right] =\tau +\frac{\left(\lambda \tau\right)^2}{2\lambda\left(1-\lambda \tau \right)},
\end{aligned}
\end{equation}
where $\lambda$ and $\tau $ are the arrival rate and the service time of the queue, respectively\footnote{{\color{black}{The entire service time of each packet is referred to as the time interval between when a packet is generated at the MTCD and when its last bit is successfully received at the BS. Because of the power constraint of our EH aided MTCD, it has a relatively low transmission rate. Therefore, the transmission duration of the packets cannot be ignored here.}}}. 
\end{lemma}
{\color{black}{\begin{Proof}
    See Section 2.1.1 in \cite{tian2006vacation}.{\hfill $\blacksquare$\par}
\end{Proof}\par%
}}
For the \textbf{MV policy}, let $\varPi_{v,n}^{0}$ denote the number of the packets generated during a single idle period for the $n$th MTCD, while the probability distribution of $\varPi_{v,n}^{0} $ can be expressed as
\begin{equation}\label{eq:L0}
  \begin{aligned}
    {\rm{Prob}}\left\{ \varPi_{v,n}^{0} =m \right\} =\frac{\left( \lambda _n\tau _s^{TM} \right) ^me^{-\lambda _n\tau _s^{TM}}}{m!}.
\end{aligned}
\end{equation}
Let $\varPi_{v,n}$ denote the number of packets generated at the beginning of a time slot. Furthermore, the probability distribution of $\varPi_{v,n}$ can be formulated as:
\begin{equation}\label{eq:L2}
  \begin{aligned}
    \rm{Prob}\left\{ \varPi_{v,n}=m \right\}&=\frac{\left( \lambda _n\tau _s^{TM} \right) ^me^{-\lambda _n\tau _s^{TM}}}{m!\left[ 1-{\rm{Prob}}\left( \varPi_{v,n}^{0} =0 \right) \right]}.
\end{aligned}
\end{equation}
where ${\rm{Prob}}\left( \varPi_{v,n}^{0} =0 \right) =e^{-\lambda _n\tau _s}$ implies the probability that there is no packet generated during one idle period and $m!=m(m-1)\cdots (1)$. Moreover, the p.g.f. of $\varPi_{v,n}$ can be expressed as follows:
\begin{equation}\label{eq:PGF-2}
  \begin{aligned}
    \varPi_{v,n}\left( z \right) &=\frac{e^{-\lambda _n\tau _s^{TM}}}{\left( 1-e^{-\lambda _n\tau _s^{TM}} \right)}\sum_{m=1}^{+\infty}{\frac{\left( \lambda _n\tau _s^{TM} \right) ^m}{m!}z^m}=\frac{e^{\lambda _n\tau _s^{TM} z}-1}{e^{\lambda _n\tau _s^{TM}}-1}.
\end{aligned}
\end{equation}
{\color{black}{Then, according to the proof of Lemma \ref{lemma:MD1-mv}, the probability generating function (p.g.f.) of the average number of packets in M/D/1 queue \textbf{without MV policy} is formulated as
\begin{equation}\label{eq:pgf1}
  \begin{aligned}
    \varPi_0^{TM} \left( z \right)=\frac{\left( 1-\lambda_n{\tau_b^{TM}}  \right) \left( 1-z \right) e^{-\left( \lambda_n-\lambda_n z \right) {\tau_b^{TM}} }}{e^{-\left( \lambda_n-\lambda_n z \right) {\tau_b^{TM}} }-z}.
  \end{aligned}
\end{equation}
Hence, we can determine the average number of packets by using the first-order derivative of $\varPi_0^{TM} \left( z \right)$. According to [1], the average number of packets \textbf{associated with the MV policy} can be derived as the sum of two independent stochastic variables, i.e., that of the number of packets with and without idle period. Therefore, the p.g.f. of the average number of packets \textbf{associated with the MV policy} is given by 
\begin{equation}\label{eq:L3}
  \begin{aligned}
    \varPi_{v}^{TM}\left( z \right) =\varPi_0^{TM} \left( z \right)\cdot \frac{1-\varPi_{v,n}\left( z \right)}{\mathbb{E} \left[ \varPi_{v,n} \right]\left(1- z \right)},
\end{aligned}
\end{equation}
which indicates that the average system delay of the MV policy can be decomposed into the sum of the average system delay without the MV policy and the additional delay due to the idle mode.}}
Under the assumption of stability, the average system delay of the $n$th MTCD using TDMA can be derived by using the first-order derivative:
\begin{equation}\label{eq:delay_1}
  \begin{aligned}
    \mathbb{E} \left[ \varPi _{v}^{TM} \right] =\frac{2\lambda _n\tau _b+\lambda _n\tau _s\left( 1-\lambda _n\tau _b \right) -\left( \lambda _n\tau _b \right) ^2}{2\left( 1-\lambda _n\tau _b \right)}.
\end{aligned}
\end{equation}
{\color{black}{Based on Little's Law, we can express the average system delay of the MV policy as $\mathbb{E} \left[ D_n^{TM} \right] =\lambda _{n}^{-1}\mathbb{E} \left[ \varPi _{v}^{TM} \right]$. Upon substituting (\ref{eq:delay_1}) into Eq. (\ref{eq:PAoI}), the average interval of the status update is $\mathbb{E} \left[ G _{n}^{TM} \right]=1/\lambda_n$, while the average peak AoI of the $n$th MTCD based on TDMA using the MV policy can be formulated as
\begin{small}
\begin{equation}\label{eq:paoi-1}
  \begin{aligned}
    \mathbb{E} \left[ A_{p,n}^{TM} \right] &= \mathbb{E} \left[ D_n^{TM} \right] +\mathbb{E} \left[ G_{n}^{TM} \right] \\
    &=\underset{The\,\,age\,\,from\,\,M/D/1}{\underbrace{\tau _{b}^{TM}+\frac{\lambda _n\left( \tau _{b}^{TM} \right) ^2}{2\left( 1-\lambda _n\tau _{b}^{TM} \right)}+\lambda _{n}^{-1}}}+\underset{Additional\,\,age}{\underbrace{\frac{\tau _s}{2}}} ,
\end{aligned}
\end{equation}
\end{small}
where $\mathbb{E} \left[ A_{p,n}^{TM} \right]$ can be decomposed into the peak AoI of the M/D/1 queue and the additional age of the idle period.}} Moreover, for the sake of simplifying the expressions of the multiple access protocol based on the MV policy, we define the average peak AoI function as ${A}_p ^{MV}\left( _n ,\tau _b,\tau _s \right)=\mathbb{E} \left[ A_{p,n}^{TM} \right]$.\par
\begin{figure}[t]
  \centering
  \includegraphics[width=0.47\textwidth]{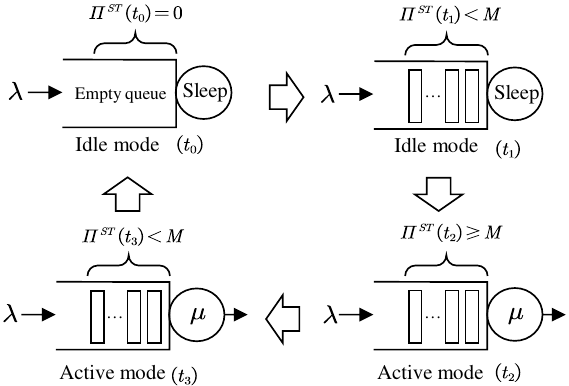}
  \caption{The State transition of the queue for one MTCD based on TDMA with the ST policy.}
  \label{fig:TDMA-NP}
  \vspace{-6mm}
\end{figure}
\textbf{(2) TDMA Using the ST Policy}: {\color{black}{\textbf{The ST policy indicates that the device in idle mode is restarted when the total number of bits generated reaches a threshold.}}} With the assistance of sleep-scheduling, the energy dissipation can be mitigated by periodically turning off the power of the MTCDs. However, frequent powering on/off degrades the energy efficiency and hence the system's lifetime. Therefore, to reduce the frequency of start-up/shutdown, we improve the TDMA protocol by introducing a start-up threshold value $M$ which determines the number of packets in the buffer, before powering up for transmission\footnote{{\color{black}{In ST policy, the switching cost is ignored since the threshold value $M$ is high enough for the extension of the idle period.}}}. \par 
Let $\varPi^{ST}\left(t\right)$ denote the number of the packets at the time instant $t$. As shown in Fig. \ref{fig:TDMA-NP}, when the queue is empty at $t=t_0$, the MTCD turns off its transmitter, switching to idle mode. Since at instant $t_1$ the number of the packets does not exceed the start-up threshold $M$, the MTCD remains in its idle mode. {\color{black}{By contract, when the number of packets exceeds $M$ at instant $t_2$, the MTCD switches to its active mode and transmits at a rate of $\mu=1/\tau_b^{ST}$ until the queue becomes empty again, where $\tau_b^{ST}$ denotes the duration of each TDMA time slot.}} Let us denote the number of packets at the end of idle mode in the MTCD by $\varPi_b^{ST}$, while its p.g.f. is given by
\begin{equation}\label{eq:TDMA-ST-1}
  \begin{aligned}
    \varPi_b^{ST}(z)&=\sum_{m=0}^{+\infty}{\rm{Prob}} \left(\varPi_b^{ST}=m\right) z^m = z^M,
\end{aligned}
\end{equation}
where we have $\mathbb{E} \left[ \varPi_b^{ST} \right]=M$. Let $\varPi_n^{{ST}{\left[ i \right]}}$ be the number of the packets in the $n$th MTCD at the $i$th packet's departure. Since the generation of the packets obeys a Poissonian process of rate $\lambda_n$, we define $\{{\varPi_n^{ST}}^{{[i]}},i\geq 1\}$ as an embedded Markov chain, which can be characterized by
\begin{equation}\label{eq:TDMA-ST-2}
  \begin{aligned}
    {\varPi _{n}^{ST}}^{\left[ i+1 \right]}=\max \left( {\varPi _{n}^{ST}}^{\left[ i \right]},0 \right) +M \mathbbm{1}_{\left\{ {\varPi _{n}^{ST}}^{\left[ i \right]}\leqslant 0 \right\}}-1+\theta^{\left[ i \right]} ,
\end{aligned}
\end{equation}
where $\theta^{\left[ i \right]}$ denotes the number of packets generated between the $i$th and the $(i+1)$st departure instant. Then, we can express the transition probability matrix of the Markov chain as follows:
\begin{equation}\label{eq:TDMA-ST-3}
  \begin{aligned}
\boldsymbol{P}_n^{ST}=\left[ \begin{matrix}
	0&		0&		\cdots&		\theta _0&		\theta _1&		\cdots\\
	\theta _0&		\theta _1&		\cdots&		\theta _{M-1}&		\theta _M&		\cdots\\
	&		\theta _0&		\cdots&		\theta _{M-2}&		\theta _{M-1}&		\cdots\\
	\boldsymbol{0}&		&		\ddots&		\ddots&		\ddots&		\ddots\\
\end{matrix} \right] ,
\end{aligned}
\end{equation}
where $\theta_m =e^{-\lambda _n\tau _b^{ST}}\frac{\left( \lambda _n\tau _b^{ST} \right) ^m}{m!},(\forall m\geqslant 0)$ represents the distribution of $\theta^{\left[ i \right]}$. We omit the index $[i]$ of $\theta_m$ because the time of transmission and the interval of update generation are independent of each other. According to Lemma \ref{lemma:MD2-ST} and Eq. (\ref{eq:PAoI}), we can derive the corresponding system delay as $\mathbb{E} \left[ D_n^{ST} \right]=\lambda _{n}^{-1}\mathbb{E} \left[ \varPi_n^{ST} \right]$, and the average interval of the status update is $\mathbb{E} \left[ G _{n}^{ST} \right]=1/\lambda_n$. {\color{black}{Similar to Eq. (\ref{eq:paoi-1}), the average peak AoI of the $n$th MTCD using TDMA having the start-up threshold $M$ can be expressed as
\begin{equation}\label{eq:paoi-2}
  \begin{aligned}
    \mathbb{E} \left[ A_{p,n}^{ST} \right] &= \mathbb{E} \left[ D_n^{ST} \right] +\mathbb{E} \left[ G_{n}^{ST} \right] \\
    &=\underset{The\,\,age\,\,from\,\,M/D/1}{\underbrace{\tau _{b}^{ST}+\frac{\lambda _n\left( \tau _{b}^{ST} \right) ^2}{2\left( 1-\lambda _n\tau _{b}^{ST} \right)}+\lambda _{n}^{-1}}}+\underset{Additional\,\,age}{\underbrace{\frac{M-1}{2\lambda _n}}},
\end{aligned}
\end{equation}
where the additional age is generated by the ST policy effect.}} In order to simplify the expressions of the multiple access protocol relying on this start-up policy, we define the average peak AoI function as ${A}_p ^{ST}\left( \lambda_n ,\tau _b^{ST}, M \right)=\mathbb{E} \left[ A_{p,n}^{ST} \right]$.
\begin{lemma}\label{lemma:MD2-ST}
When the queue is stable, i.e., $\rho=\lambda\tau_n<1$, the average number of packets generated by the $n$th MTCD based on TDMA using the start-up threshold $M$ is given by
\begin{equation}\label{eq:lemma2-1}
  \begin{aligned}
    \mathbb{E} \left[ \varPi_n^{ST} \right] =\lambda _n\tau _b+\frac{\left( \lambda _n\tau _b \right) ^2}{2\left( 1-\lambda _n\tau _b \right)}+\frac{M-1}{2},
\end{aligned}
\end{equation}
where $\lambda_n$ and $\tau_n^{-1}$ denote the rates of update generation and of the packet transmission, respectively.
\end{lemma}
\begin{Proof}
    See Appendix \ref{proof:lemma-2}.{\hfill $\blacksquare$\par}
\end{Proof}\par
\subsection{FDMA Protocol}
\label{sec:PAoI_FDMA}
\begin{figure}[t]
  \centering
  \includegraphics[width=0.47\textwidth]{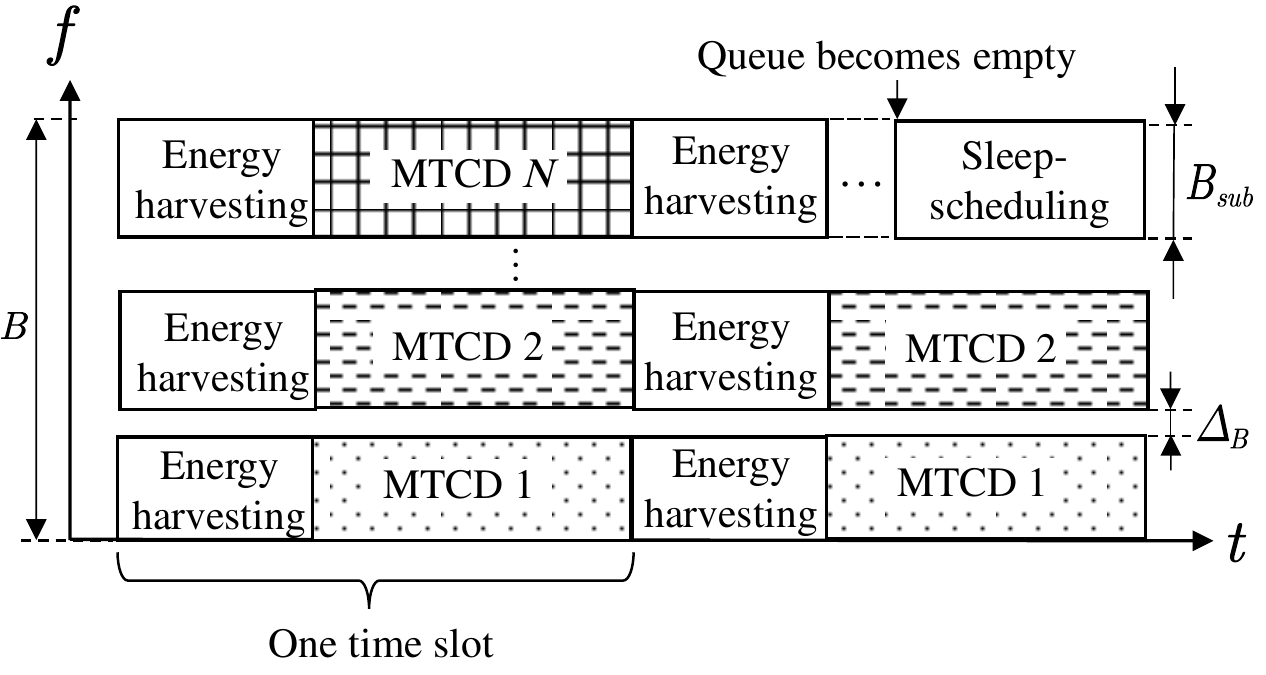}
  \caption{An example that illustrates the FDMA with sleep-scheduling.}
  \label{fig:FDMA-1}
  \vspace{-6mm}
\end{figure}
In FDMA, the bandwidth $B$ has been divided into orthogonal frequency subchannels, and each sub-channel is allocated to a single MTCD. By sharing the same channel, all MTCDs send their updates generated simultaneously but in different frequency bands. As shown in Fig. \ref{fig:FDMA-1}, we assume that all MTCDs have the same bandwidth allocation $B_{Sub}$ and the guard band\footnote{The guard band of the FDMA protocol represents the unused part of the radio spectrum between each sub-channel, which avoids the adjacent channel interference.} is $\varDelta_B$. Furthermore, we have $B_{Sub}=N^{-1}\left[ B-\left( N-1 \right) \varDelta _B \right] $. To differentiate our policies, we use superscripts $FM$ and $FS$ to denote the FDMA MV policy and ST policy, respectively.

\textbf{(1) FDMA Using the MV Policy}: The analysis of the average peak AoI for FDMA using the MV policy is similar to that in Section \ref{sec:PAoI_TDMA-IB}. In contrast to TDMA, the entire transmission time $\tau_{b}^{FM}$ of a sub-channel is used for a single MTCD, but its bandwidth is only $B_{Sub}\approx \frac{B}{N}$. Furthermore, let $\tau_{s}^{FM}$ and ${\tau_{p}^{FM}}=\tau_p$ represent the durations of the sleep-scheduling and EH within an FDMA time slot. According to Lemma \ref{lemma:MD1-mv} and Eq. (\ref{eq:PAoI}), we have an average peak AoI for the $n$th MTCD based on the FDMA using the MV policy given by
\begin{equation}\label{eq:paoi-3}
  \begin{aligned}
    \mathbb{E} \left[ A_{p,n}^{FM} \right] &={A}_p ^{MV}\left( \lambda_n ,\tau_{b}^{FM},{\tau_{s}^{FM}} \right),
\end{aligned}
\end{equation}
where the definition of the function ${A}_p ^{MV}\left( \lambda_n ,\tau_{b}^{FM},{\tau_{s}^{FM}} \right)$ is obtained according to Eq. (\ref{eq:paoi-1}).

\textbf{(2) FDMA Using the ST Policy}: For the sake of avoiding frequently turning on/off the MTCDs, we improve the conventional FDMA by introducing a start-up threshold $M$. Specifically, based on the features of FDMA using multiple idle TSs, the transmissions are only triggered, when the status update counter has reached $M$. Then, the MTCD continues its transmissions until the MTCD queue becomes empty. The procedure is visualized in Fig. \ref{fig:TDMA-NP}. Similar to the definitions of Section \ref{sec:PAoI_FDMA}, $\tau_{b}^{FS}$ and ${\tau_{s}^{FS}}$ represent the duration of the active and idle mode, respectively. Inspired by the proof of Lemma \ref{lemma:MD2-ST}, we can express the average of the peak AoI as follows:
\begin{equation}\label{eq:paoi-4}
  \begin{aligned}
    \mathbb{E} \left[ A_{p,n}^{FS} \right] &={A}_p ^{ST}\left( \lambda_n ,\tau _b^{FS}, M \right),
\end{aligned}
\end{equation}
where the definition of the function ${A}_p ^{ST}\left( \lambda_n ,\tau _b^{FS}, M \right)$ is obtained according to Eq. (\ref{eq:paoi-2}).
\subsection{NOMA Protocol}
\label{sec:PAoI_NOMA}
\begin{figure}[t]
  \centering
  \includegraphics[width=0.47\textwidth]{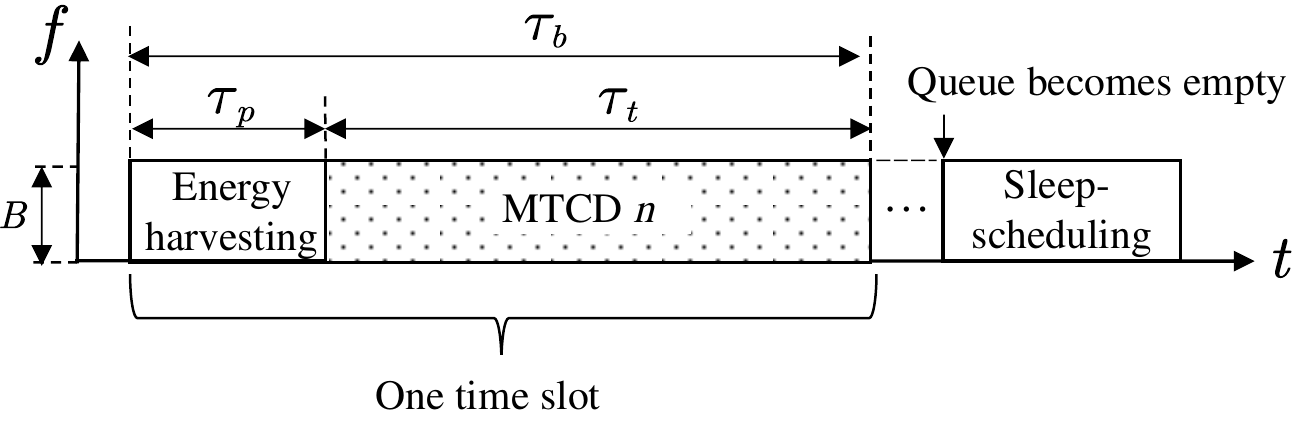}
  \caption{An example that illustrates the NOMA with sleep-scheduling.}
  \label{fig:NOMA-1}
  \vspace{-6mm}
\end{figure}

\textbf{(1) NOMA Using the MV Policy}: In power-domain NOMA (PD-NOMA), all MTCDs share the same resources both in terms of time, frequency, and space. As shown in Fig. \ref{fig:NOMA-1}, we assume that $\tau_{b,n}^{NM}=\tau_{b}$ denotes the duration of a NOMA time slot partitioned into one EH TS and one transmission TS in the $n$th MTCD, where $\tau_{p,n}^{NM}=\tau_{p}$ and $\tau_{t,n}^{NM}=\tau_{b}-\tau_{p}$ are the duration of EH and transmission within a time slot, respectively.
According to Lemma \ref{lemma:MD1-mv} and Eq. (\ref{eq:PAoI}), the average peak AoI of the EH-aided multiple access network can be formulated as:
\begin{small}
\begin{equation}\label{eq:noma-5}
  \begin{aligned}
    \mathbb{E} \left[ A_{p}^{NM} \right] &=\frac{1}{N}\sum_{n=1}^N{{A}_p ^{MV}\left( \lambda_n ,\tau _{b}^{NM},\tau _{s}^{NM} \right)}={{A}_p ^{MV}\left( \lambda_n ,\tau _{b}^{NM},\tau _{s}^{NM} \right)}.
\end{aligned}
\end{equation}
\end{small}
\par
\textbf{(2) NOMA Using the ST Policy}: In order to reduce the frequency of turning on/off the MTCDs, we use the ST policy for improving the start-up threshold. According to the above derivation of the different protocols based on the ST policy of Sections \ref{sec:PAoI_TDMA-IB} and \ref{sec:PAoI_FDMA}, we can express the average peak AoI of the NOMA using the ST policy as
\begin{equation}\label{eq:paoi-6}
  \begin{aligned}
    \mathbb{E} \left[ A_{p,n}^{NS} \right] &=\frac{1}{N}\sum_{n=1}^N{A}_p ^{NS}\left( \lambda_n ,\tau _{b,n}^{NS}, M \right)={A}_p ^{NS}\left( \lambda_n ,\tau _{b}^{NS}, M \right),
\end{aligned}
\end{equation}
where $\tau _{b,n}^{NS}$ and $M$ denote the duration of one NOMA time slot and the start-up threshold.
\section{Problem Formulation}
\label{Problem Formulation}
In this section, based on TDMA, FDMA and PD-NOMA relying on different policies, we will first formulate the problems of minimizing the average peak AoI, which can be transformed into convex problems, and their optimal solutions are obtained via an exact linear search method (Algorithm \ref{alg:ELS}). The original non-convex problems are formulated as DC programming problems, which are solved by the CCP method (Algorithm \ref{alg:CCP}) for reducing their computational complexity. 
\subsection{TDMA-based Problem Formulation}
\label{sec:PF-TDMA}
\textbf{(1) TDMA Using MV Policy}: Without loss of generality, we assume that every MTCD has the same protocol parameters, namely $\tau_{p,n}=\tau_{p}$, $\tau_{t,n}=\tau_{t}$, $\tau_{s,n}=\tau_{s}$, $(n=1,2,\cdots,N)$. Moreover, the energy efficiency (EE) of the status updates scheme can be defined as the ratio of the generation rate and the sum of the transmit power consumption\cite{tang2019energy}. According to \ref{eq:capacity TDMA}, the transmit power of the $n$th MTCD during the selected TS is given by $\varPhi _{t,n}^{TM}=\left[ 2^{\frac{\left( \mathcal{J} +1 \right)NL}{\left( \tau _{b}^{TM}-\tau _{p}^{TM} \right) {B}}}-1 \right] \gamma _n^{-1}$. Therefore, the EE of the considered status update scheme based on TDMA using the MV policy can be formulated as:
\begin{equation}\label{eq:tdma-1-ee}
  \begin{aligned}
\varUpsilon _{n}^{TM}&=\frac{NL}{\left( \tau _{b}^{TM}-\tau _{p}^{TM} \right) \varPhi _{t,n}^{TM}}=\frac{\gamma _n NL}{\left( \tau _{b}^{TM}-\tau _{p}^{TM} \right)\left[ 2^{\frac{\left( \mathcal{J} +1 \right)NL}{\left( \tau _{b}^{TM}-\tau _{p}^{TM} \right) {B}}}-1 \right]} .
\end{aligned}
\end{equation}
Furthermore, we assume that the EE of the system should not be lower than $\varUpsilon _{min}$, namely $\varUpsilon _{n}^{TM}\geq\varUpsilon _{min}$ where leads to the following inequality:
\begin{equation}\label{eq:tdma-1-ee2}
  \begin{aligned}
    \varUpsilon _{\min}\left( \tau _{b}^{TM}-\tau _{p}^{TM} \right)\left[ 2^{\frac{\left( \mathcal{J} +1 \right)NL}{\left( \tau _{b}^{TM}-\tau _{p}^{TM} \right) {B}}}-1 \right] \leqslant NL\gamma _n,
\end{aligned}
\end{equation}
where the EH time $\tau _{p}^{TM}$ is a constant. When optimizing the average peak AoI and the EE of the system, the associated minimization problem $ \mathbf{P}_{1}^{TM}$ is expressed as
\begin{equation}\label{OP:ET-1-1}
  \begin{aligned}
    \mathbf{P}_{1}^{TM}: \quad \!&\min \limits_{\boldsymbol{x}^{TM}} {A}_p ^{MV}\left( \lambda_n ,\tau _b^{TM},\tau _s^{TM} \right)\\
    \textrm{s.t.} \quad
    (\ref{OP:ET-1-1}\textrm{a}):& \quad\left(\frac{\tau _{b}^{TM}-\tau _{p}^{TM}}{N}  \right) \varPhi _{t,n}^{TM}\le \tau _{p}^{TM} \varPhi _{r,n}\le E_{b,\max},\\
    (\ref{OP:ET-1-1}\textrm{b}):& \quad \left[ \tau _{p}^{TM},0 \right]\preceq \boldsymbol{\tau }^{TM}\preceq \left[ \tau _{b,\max}^{TM},\tau _{s,\max}^{TM} \right] ,\\
    (\ref{OP:ET-1-1}\textrm{c}):& \quad\lambda _{n,\min}\leqslant \lambda _n\leqslant \lambda _{n,\max},\\
    (\ref{OP:ET-1-1}\textrm{d}):& \quad0<\lambda _n\tau _{b}^{TM}<1\ \rm{and}\ (\ref{eq:tdma-1-ee2}),
  \end{aligned}
\end{equation}
\begin{algorithm}[!t]
  \caption{Age-optimal Exact Linear Search}
  \begin{algorithmic}[1]
  \STATE Constrain $\lambda_n$ $\lambda _{n,\min}\leqslant \lambda _n\leqslant \min \left( \lambda _{n,\max},1/\tau _{b}^{TM} \right). $
  \STATE With the fixed $\lambda_{n}$, the original problem is converted to a convex sub-problem.
  \STATE Initialize the starting point to $\lambda_{(0)}=\lambda _{n,\min}$, the iteration threshold to $K$, and the iteration indicator to $k:=0$.
  \STATE Set the accuracy (step size), to ${A} =\left[ \min \left( \lambda _{n,\max},1/\tau _{b}^{TM} \right) -\lambda _{n,\min} \right] /K.$
  \REPEAT
  \STATE Solve the convex sub-problem with the solution $\boldsymbol{x}_{\left( k \right)}$, and find the optimal solution $\boldsymbol{x}_{(*)}=\max  \boldsymbol{x}_{\left( i \right)},\forall i\leqslant k$.
  \STATE Update $\lambda_{(k+1)}=\lambda_{(k)}+{A}$ and $k:=k+1$.
  \UNTIL $k>K$.
  \STATE \textbf{Output} Optimal solution $\boldsymbol{x}_{(*)}$.
  \end{algorithmic}\label{alg:ELS}
\end{algorithm}
where the objective function is the average peak AoI ${A}_p ^{MV}\left( \lambda_n ,\tau _b^{TM},\tau _s^{TM} \right)$, while $\boldsymbol{x}^{TM}=\left\{ \lambda_n,\tau _{b}^{TM},\tau _{s}^{TM},\varPhi _{r,n} \right\} $ denotes the vector of the associated variables and $\boldsymbol{\tau }^{TM}=\left\{\tau _{b}^{TM},\tau _{s}^{TM}\right\}$ is the vector of TS allocation. Furthermore, $\lambda_n$ and $\varPhi _{r,n}^{TM}$ denote the rate of status updates and the transmit power of the power station, respectively. The inequality constraint in (\ref{OP:ET-1-1}a) implies that the energy harvested should be higher than the transmit power consumed but less than or equal to the battery capacity $E_{b,\max}$. Furthermore, (\ref{OP:ET-1-1}b) and (\ref{OP:ET-1-1}c) denote the inequality constraints of the slot allocation and the generation rate of status updates for the $n$th MTCD, respectively. Moreover, the optimization problem should yield the stable condition in $(\ref{OP:ET-1-1}d)$. According to Lemma \ref{lemma:4}, we first convert the non-convex problem in (\ref{OP:ET-1-1}) to a convex one by fixing the status update rate $\lambda_n$. Then, we use the exact line search based method to find the optimal solution, while the main procedure of the search is shown in Algorithm \ref{alg:ELS}.

\begin{lemma}\label{lemma:4}
Consider the minimization problem in (\ref{OP:ET-1-1}). For the sake of obtaining the optimal solution, we fix the variable $\lambda_n$ to as a constant $\lambda_{(k)}$, where $\lambda_{(k)}$ yields the constraint of $\lambda_n$. According to the second-order condition, the original objective function is converted to a convex function ${\widehat{{A}}} _{p}^{MV}\left( \tau _{b}^{TM},\tau _{s}^{TM} \right) \stackrel{\triangle}{=}{A} _{p}^{MV}\left( \lambda _{\left( \mathrm{k} \right)},\tau _{b}^{TM},\tau _{s}^{TM} \right) $. Similarly, the objective function $ {A}_p ^{ST}\left( \lambda_n ,\tau _b^{ST},M \right)$ in (\ref{OP:ST-1}) can be converted to a convex function $ {\widehat{{A}}} _{p}^{ST}\left( \tau _{b}^{TM},M \right)$ with the aid of the same operation.
\end{lemma}
\begin{Proof}
  We assume that ${\widehat{{A}}} _{p}^{MV}\left( \tau _{b}^{TM},\tau _{s}^{TM} \right)$ and $ {\widehat{{A}}} _{p}^{ST}\left( \tau _{b}^{TM},M \right)$ are twice differentiable. Then we can derive their Hessian matrices as $\nabla ^2\widehat{{A} }_{p}^{MV}=\mathrm{diag}\left[ \lambda _n\left( 1-\lambda _n\tau _{b}^{TM} \right) ^{-3},0 \right]  \succeq 0$ and $\nabla ^2\widehat{{A} }_{p}^{ST}=\mathrm{diag}\left[ \lambda _n\left( 1-\lambda _n\tau _{b}^{TM} \right) ^{-3},0 \right]  \succeq 0$. Therefore, the optimization problems in (\ref{OP:ET-1-1}) and (\ref{OP:ST-1}) can be converted to convex subproblems with a fixed $\lambda_n$. 
\end{Proof}\par
\begin{lemma}\label{lemma:3}
  The average peak AoI function of the system using the MV policy ${A}_p ^{MV}\left( \lambda_n ,\tau _b^{TM},\tau _s^{TM} \right)$ in Eq. (\ref{eq:paoi-1}) is the sum of the convex function $f_{1}^{TM}\left( \boldsymbol{x}^{TM} \right)$ and of the concave function $f_{2}^{TM}\left( \boldsymbol{x}^{TM} \right)$. Thus, the original problem $ \mathbf{P}_{1}^{TM}$ can be converted to a DC programming problem:
  \begin{equation}\label{OP:ET-1-2}
    \begin{aligned}
      \mathbf{P}_{2}^{TM}: \quad \!&\min \limits_{\boldsymbol{x}^{TM}} f_{1}^{TM}\left( \boldsymbol{x}^{TM} \right) +\boldsymbol{x}^{TM}\nabla f_{2}^{TM}\left( {\boldsymbol{x}^{TM}_{(k)}} \right)^\mathrm{T}\\
      \textrm{s.t.} \quad
      & (\ref{eq:tdma-1-ee2}),\ (\ref{OP:ET-1-1}\textrm{a}),\ (\ref{OP:ET-1-1}\textrm{b}),\ (\ref{OP:ET-1-1}\textrm{c}),\\
      & \left[ \lambda _n\left( \tau _{b}^{TM\,\,} \right) _{\left( k \right)}+\tau _{b}^{TM\,\,}\left( \lambda _{n}^{\,\,} \right) _{\left( k \right)} \right] <1+\left( \lambda _n\tau _{b}^{TM\,\,} \right) _{\left( k \right)},
    \end{aligned}
  \end{equation}
  where ${\boldsymbol{x}^{TM}_{(k)}}$ denotes the optimal solution obtained from the $k$th iteration, while $\nabla f_{2}^{TM}\left( {\boldsymbol{x}^{TM}_{(k)}} \right)^\mathrm{T}$ is the gradient value of $ f_{2}^{TM}\left( {\boldsymbol{x}^{TM}_{(k)}} \right)$ at the point ${\boldsymbol{x}^{TM}_{(k)}}$. After successively solving a series of convex problems, we can find a locally optimal solution for the minimization problem $\mathbf{P}_{1}^{TM}$.
  \end{lemma}
  \begin{Proof}
      See Appendix \ref{proof:lemma-3}.{\hfill $\blacksquare$\par}
  \end{Proof}
  
\begin{algorithm}[!t]
  \caption{CCP Based Peak AoI Optimization}
  \begin{algorithmic}[1]
  \STATE Initialize the feasible starting point $\boldsymbol{x}_{(0)}^{TM}$, the iteration step threshold $K$ and constant $\epsilon$.
  \STATE Initialize the convex function $f_{1}^{TM}(\boldsymbol{x})$ and the concave function $f_{2}^{TM}(\boldsymbol{x})$ by Lemma \ref{lemma:3}.
  \STATE Set the lower and the upper bounds of $\lambda_n$, $\tau_b$, $\tau_s$, $\varPhi _{r,n}$.
  \STATE Set the iteration indicator to $k:=0$.
  \REPEAT
  \STATE Calculate the gradient value $\nabla f_{2}^{TM}$ and substitute a given $\boldsymbol{x}_{(k)}^{TM}$. 
  \STATE Obtain the convex sub-problem in (\ref{OP:ET-1-2}).
  \STATE Calculate the locally optimal solution of (\ref{OP:ET-1-2}) as $\boldsymbol{x}_{(*)}^{TM}$.
  \STATE Update $\boldsymbol{x}_{(k+1)}^{TM}=\boldsymbol{x}_{(*)}^{TM}$ and $k:=k+1$.
  \UNTIL $k>K$ or $\left| \boldsymbol{x}_{(k+1)}^{TM}-\boldsymbol{x}_{(k)}^{TM} \right|<\epsilon $.
  \STATE \textbf{Output} near-optimal solution $\boldsymbol{x}_{(*)}^{TM}$.
  \end{algorithmic}\label{alg:CCP}
\end{algorithm}
To increase the accuracy of the solution and hence to approach the optimum, increasing complexity of Algorithm \ref{alg:ELS} is inevitable, which is counterproductive for EH-aided MTCDs having finite computational resources. Therefore, we propose a low-complexity heuristic algorithm for solving the original problem. According to Lemma \ref{lemma:3}, the objective function of the minimization problem $ \mathbf{P}_{1}^{TM}$ is non-convex, while $ \mathbf{P}_{1}^{TM}$ belongs to the family of DC programming problems, which can be solved by the CCP algorithm. {\color{black}{Specifically, we decompose the original objective function ${A}_p ^{MV}\left( \boldsymbol{x}^{TM} \right)$ into the sum of a convex and a concave function, i.e. $\frac{\tau _{s}^{TM}}{2}+\frac{\tau _{b}^{TM}}{2}+\frac{1}{\lambda _n}$ and $\frac{\tau _{b}^{TM}}{2\left( 1-\lambda _n\tau _{b}^{TM} \right)}$. Then, we determine an initial feasible point $\boldsymbol{x}^{TM}_0$ and convert ${A}_p ^{MV}\left( \boldsymbol{x}^{TM}\right)$ to typical convex function relying on first order Taylor approximation as follows:
\begin{small}
\begin{equation}\label{eq:CCP_taylor}
  \begin{aligned}
    \widehat{{A} }_{p}^{MV}\left( \boldsymbol{x}^{TM};\boldsymbol{x}_{0}^{TM} \right)&=\widehat{{A} }_{p}^{MV}\left( \boldsymbol{x}_{0}^{TM} \right)\\
    &+\nabla \left[ \widehat{{A} }_{p}^{MV}\left( \boldsymbol{x}_{0}^{TM} \right) \right] ^{\mathrm{T}}\left( \boldsymbol{x}^{TM}-\boldsymbol{x}_{0}^{TM} \right). 
\end{aligned}
\end{equation}
\end{small}
Then, substituting ${A}_p ^{MV}\left( \boldsymbol{x}^{TM} \right)$ by $\widehat{{A} }_{p}^{MV}\left( \boldsymbol{x}^{TM};\boldsymbol{x}_{0}^{TM} \right)$ in $\mathbf{P}_{1}^{TM}$ and $\mathbf{P}_{1}^{TM}$ can be converted to
a classical convex problem $\mathbf{P}_{2}^{TM}$. During each iteration, $\mathbf{P}_{2}^{TM}$ is solved for a given $\boldsymbol{x}_{0}^{TM}$, which can be replaced by the optimal value in the next iteration. }}The algorithm stops the iteration when the stopping criterion is satisfied. The main steps of the CCP algorithm are shown in Algorithm \ref{alg:CCP}.
\par
\begin{figure}[t]
  \centering
  \includegraphics[width=0.52\textwidth]{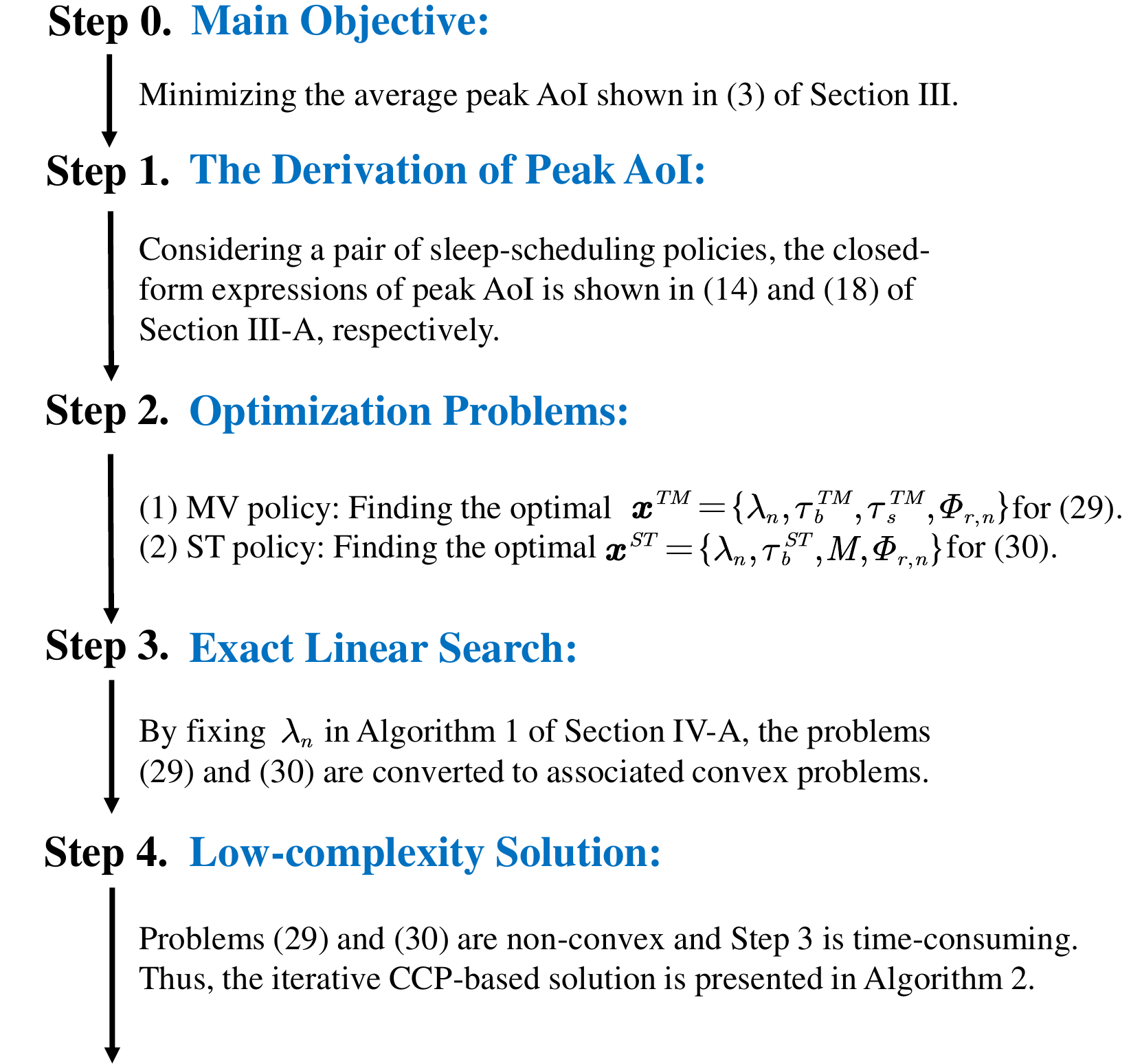}
  \caption{Flow of the mathematical analysis for TDMA protocol.}
  \label{fig:flow}
  \vspace{-6mm}
\end{figure}

\textbf{(2) TDMA using the ST Policy}: The EE of the status update scheme based on TDMA and the ST policy is given by $\varUpsilon _{n}^{ST}=\frac{L}{(\tau _{b}^{ST}-\tau _{p}^{ST}) \varPhi _{t,n}^{ST}}$. Similar to Section \ref{sec:PF-TDMA}(1), the lower bound of the EE yields $\varUpsilon _{n}^{ST}\geq\varUpsilon _{\min}$ and the minimization problem $ \mathbf{P}_{1}^{ST}$ is expressed as
\begin{equation}\label{OP:ST-1}
  \begin{aligned}
    \mathbf{P}_{1}^{ST}: \quad \!&\min \limits_{\boldsymbol{x}^{ST}} {A}_p ^{ST}\left( \lambda_n ,\tau _b^{ST},M \right)\\
    \textrm{s.t.} \quad
    (\ref{OP:ST-1}\textrm{a}):& \quad\left( \frac{\tau _{b}^{ST}-\tau _{p}^{ST}}{N} \right) \varPhi _{t,n}^{ST}\le \tau _{p}^{ST} \varPhi _{r,n}\le E_{b,\max},\\
    (\ref{OP:ST-1}\textrm{b}):& \quad \tau _{p}^{ST}\leq\tau _{b}^{ST}\leq   \tau _{b,\max}^{ST} ,\\
    (\ref{OP:ST-1}\textrm{c}):& \quad\lambda _{n,\min}\leqslant \lambda _n\leqslant \lambda _{n,\max},\\
    (\ref{OP:ST-1}\textrm{d}):& \quad0<\lambda _n\tau _{b}^{ST}<1,\\
    (\ref{OP:ST-1}\textrm{e}):& \quad\varUpsilon _{n}^{ST}\geq\varUpsilon _{min},
  \end{aligned}
\end{equation}
where $\boldsymbol{x}^{ST}=\left\{ \lambda_n,\tau _{b}^{ST},M,\varPhi _{r,n} \right\} $ denotes the vector of the variables. The optimal solution of problem (\ref{OP:ST-1}) can be found by Algorithm \ref{alg:ELS}. In order to simplify the original problem, we relax the integrality constraint to transfer $M$ to a non-integral value, i.e. $M\in[M_{min},M_{max}]$. Furthermore, according to Eq. (\ref{eq:paoi-2}), the objective function of the problem $ \mathbf{P}_{1}^{ST}$ is the sum of a convex function $f_{1}^{ST}\left( \lambda _{n},\tau _{b}^{ST},M \right) =\tau _{b}^{ST}+(2\lambda _{n})^{-1}$ and of a concave function $f_{2}^{ST}\left( \lambda _{n},\tau _{b}^{ST},M \right) =\frac{\lambda _n\left( \tau _{b}^{ST} \right) ^2}{2\left( 1-\lambda _n\tau _{b}^{ST} \right)}+\frac{M}{2\lambda _n}$. Inspired by Lemma \ref{lemma:3}, $ \mathbf{P}_{1}^{ST}$ can be rewritten as a DC program as follows:
\begin{equation}\label{OP:ET-2-2}
  \begin{aligned}
    \mathbf{P}_{2}^{ST}: \quad \!&\min \limits_{\boldsymbol{x}^{ST}} f_{1}^{ST}\left( \boldsymbol{x}^{ST} \right) +\boldsymbol{x}^{ST}\nabla f_{2}^{ST}\left( {\boldsymbol{x}^{ST}_{(k)}} \right)^\mathrm{T}\\
    \textrm{s.t.} \quad
    &(\ref{OP:ST-1}\textrm{a}),\ (\ref{OP:ST-1}\textrm{b}),\ (\ref{OP:ST-1}\textrm{c}),\ (\ref{OP:ST-1}\textrm{e}),\\
    & \left[ \lambda _n\left( \tau _{b}^{ST\,\,} \right) _{\left( k \right)}+\tau _{b}^{ST\,\,}\left( \lambda _{n}^{\,\,} \right) _{\left( k \right)} \right] <1+\left( \lambda _n\tau _{b}^{ST\,\,} \right) _{\left( k \right)},
  \end{aligned}
\end{equation}
where ${\boldsymbol{x}^{ST}_{(k)}}$ denotes the optimal solution obtained from the $k$th iteration, and $ \mathbf{P}_{2}^{ST}$ can be solved by the CCP method of Algorithm \ref{alg:CCP}. Before proceeding the following sections, we depict an example diagram shown in Fig. \ref{fig:flow} to  summarize the analysis flow of TDMA\footnote{Fig. \ref{fig:flow} is not only for TDMA protocol but also for FDMA and NOMA.}.
\subsection{FDMA-based Problem Formulation}
\label{sec:PF-FDMA}
\textbf{(1) FDMA Using the MV Policy}: In FDMA, the bandwidth of the $n$th MTCD subchannel is $B_{Sub}=N^{-1}\left[ B-\left( N-1 \right) {A} _B \right]$. Thus, the EE of the FDMA using the MV policy can be formulated as $\varUpsilon _{n}^{FM}=\frac{L}{\left( \tau _{b}^{FM}-\tau _{p}^{FM} \right) \varPhi _{t,n}^{FM}}$, where the transmit power is given by $\varPhi _{t,n}^{FM}=\gamma _{n}^{-1}\left[ 2^{\frac{\left( \mathcal{J} +1 \right)L}{\left( \tau _{b}^{FM}-\tau _{p}^{FM} \right) B_{Sub}}}-1 \right]$. Therefore, the lower bound of the EE yields
\begin{equation}\label{eq:fdma-1-ee2}
  \begin{aligned}
    \varUpsilon _{\min}\left( \tau _{b}^{FM}-\tau _{p}^{FM} \right) \left[ 2^{\frac{\left( \mathcal{J} +1 \right)L}{\left( \tau _{b}^{FM}-\tau _{p}^{FM} \right) B_{Sub}}}-1 \right] \leqslant \gamma _nL.
\end{aligned}
\end{equation}
The problem of optimizing the average peak AoI and the EE of the system may be formulated as the minimization problem $ \mathbf{P}^{FM}$:
\begin{equation}\label{OP:EF-1}
  \begin{aligned}
    \mathbf{P}^{FM}: \quad \!&\min \limits_{\boldsymbol{x}^{FM}} {A}_p ^{MV}\left( \lambda_n ,\tau _b^{FM},\tau _s^{FM} \right)\\
    \textrm{s.t.} \quad
    (\ref{OP:EF-1}\textrm{a}):& \quad\left( \tau _{b}^{FM}-\tau _{p}^{FM} \right) \varPhi _{t,n}^{FM}\le \tau _{p}^{FM} \varPhi _{r,n}\le E_{b,\max},\\
    (\ref{OP:EF-1}\textrm{b}):& \quad \left[ \tau _{p}^{FM},0 \right]\preceq \boldsymbol{\tau }^{FM}\preceq \left[ \tau _{b,\max}^{FM},\tau _{s,\max}^{FM} \right] ,\\
    (\ref{OP:EF-1}\textrm{c}):& \quad\lambda _{n,\min}\leqslant \lambda _n\leqslant \lambda _{n,\max},\\
    (\ref{OP:EF-1}\textrm{d}):& \quad0<\lambda _n\tau _{b}^{FM}<1\ \rm{and}\ (\ref{eq:fdma-1-ee2}),
  \end{aligned}
\end{equation}
where the objective function is the average peak AoI, while $\boldsymbol{x}^{FM}=\left\{ \lambda_n,\tau _{b}^{FM},\tau _{s}^{FM},\varPhi _{r,n} \right\} $ denotes the vector of the variables and $\boldsymbol{\tau }^{FM}$ is the vector of TS allocation. \par
\textbf{(2) FDMA Using the ST Policy}: Under the ST policy, the MTCD only wakes up from its idle mode, if the number of status updates in the queue is more than $M$. Moreover, the EE of the system yields $\varUpsilon _{n}^{FS}=\frac{NL}{(\tau _{b}^{FS}-\tau _{p}^{FS}) \varPhi _{t,n}^{FS}}\geq\varUpsilon _{\min}$. Similarly to the minimization problem above, let $\boldsymbol{x}^{FS}=\left\{ \lambda_n,\tau _{b}^{FS},M,\varPhi _{r,n} \right\} $ denote the vector of variables, and the optimization problem is
\begin{equation}\label{OP:FST-1}
  \begin{aligned}
    \mathbf{P}^{FS}: \quad \!&\min \limits_{\boldsymbol{x}^{FS}} {A}_p ^{ST}\left( \lambda_n ,\tau _b^{FS},M \right)\\
    \textrm{s.t.} \quad
    (\ref{OP:FST-1}\textrm{a}):& \quad\left( \tau _{b}^{FS}-\tau _{p}^{FS} \right) \varPhi _{t,n}^{FS}\le \tau _{p}^{FS} \varPhi _{r,n}\le E_{b,\max},\\
    (\ref{OP:FST-1}\textrm{b}):& \quad \tau _{p,\min}^{FS}\leq\tau _{b}^{FS}\leq   \tau _{b,\max}^{FS}  ,\\
    (\ref{OP:FST-1}\textrm{c}):& \quad\lambda _{n,\min}\leqslant \lambda _n\leqslant \lambda _{n,\max},\\
    (\ref{OP:FST-1}\textrm{d}):& \quad0<\lambda _n\tau _{b}^{FS}<1,\\
    (\ref{OP:FST-1}\textrm{e}):& \quad\varUpsilon _{n}^{FS}\geq\varUpsilon _{min}.
  \end{aligned}
\end{equation}\par
According to Lemmas \ref{lemma:4} and \ref{lemma:3}, the optimal results of $\mathbf{P}^{FM}$ and $\mathbf{P}^{FS}$ can be obtained by Algorithm \ref{alg:ELS}. To reduce the computational complexity, the previous problems may be rewritten as DC programming problems. Algorithm \ref{alg:CCP} provides a tractable iterative procedure based on a CCP method to solve it. 
\subsection{NOMA-based Problem Formulation}
\label{sec:PF-NOMA}
\textbf{(1) NOMA Using the MV Policy}: For the case of NOMA, we assume that all the $N$ MTCDs share the same spectrum along with the same status update rate and TS allocation. Since SIC is utilized by the BS receiver, we should accurately control the transmit power and the EH of the different MTCDs. The EE of the NOMA scheme using the MV policy is the ratio between the status update rate and the average power consumption, expressed as: 
\begin{equation}\label{OP:EN-1-ee}
  \begin{aligned}
    \varUpsilon ^{NM}=\frac{\gamma LN}{\left( \tau _{b}^{NM}-\tau _{p}^{NM} \right) \left[ 2^{\frac{\left( \mathcal{J} +1 \right)NL}{\left( \tau _{b}^{NM}-\tau _{p}^{NM} \right) B}}-1 \right]}\geq\varUpsilon _{\min}.
  \end{aligned}
\end{equation}
In contrast to the TDMA and FDMA protocols, the transmit power of NOMA should satisfy the power constraint of Eq. (\ref{eq:noma-2}). Therefore, the corresponding minimization problem is
\begin{equation}\label{OP:EN-1}
  \begin{aligned}
    \mathbf{P}^{NM}: \quad \!&\min \limits_{\boldsymbol{x}^{NM}} {A}_p ^{MV}\left( \lambda ,\tau _b^{NM},\tau _s^{NM} \right)\\
    \textrm{s.t.} \quad
    (\ref{OP:EN-1}\textrm{a}):& \quad\left( \tau _{b}^{NM}-\tau _{p}^{NM} \right) P_{\Sigma }\le \tau _{p}^{NM} \varPhi _{\Sigma }\le E_{b,\max},\\
    (\ref{OP:EN-1}\textrm{b}):& \quad \left[ \tau _{p}^{NM},0 \right]\preceq \boldsymbol{\tau }^{NM}\preceq \left[ \tau _{b,\max}^{NM},\tau _{s,\max}^{NM} \right] ,\\
    (\ref{OP:EN-1}\textrm{c}):& \quad\lambda _{\min}\leqslant \lambda \leqslant \lambda _{\max},\\
    (\ref{OP:EN-1}\textrm{d}):& \quad0<\lambda \tau _{b}^{NM}<1,\\
    (\ref{OP:EN-1}\textrm{f}):& \quad \varUpsilon ^{NM}\geq\varUpsilon _{\min},\ {\rm{and}} \ (\ref{OP:EN-1-ee}),
  \end{aligned}
\end{equation}
where $(\ref{OP:EN-1}\textrm{a})$ is the transmit power constraint of NOMA, and we have $P _{\Sigma }=\sum_{n=1}^N{\varPhi _{t,n}^{NM}}=\left[ 2^{\frac{\left( \mathcal{J} +1 \right)NL}{\left( \tau _{b}^{NM}-\tau _{p}^{NM} \right) B}}-1 \right] /\gamma $, where $\varPhi _{t,n}^{NM}\geq 0, \forall {n}\in \mathcal{N}$. Let $\boldsymbol{x}^{NM}=\left\{ \lambda,\tau _{b}^{NM},\tau _{s}^{NM},\varPhi _{\Sigma } \right\}$ denote the vector of the variables. Moreover, we omit the subscript $n$ since all MTCDs have the same average peak AoI along with the same TS allocation and status update rate. 
\par
\textbf{(2) NOMA Using the ST Policy}: In this policy, the optimization problem requires the minimization of the average peak AoI ${A}_p ^{ST}\left( \lambda_n ,\tau _b^{NS},M \right)$ under the constraints of transmit power, TS allocation etc. Thus, the problem for the case of the NOMA using the ST policy can be formulated as follows, while we drop the subscript $n$ for simplicity
\begin{equation}\label{OP:NST-1}
  \begin{aligned}
    \mathbf{P}^{NS}: \quad \!&\min \limits_{\boldsymbol{x}^{NS}} {A}_p ^{ST}\left( \lambda_n ,\tau _b^{NS},M \right)\\
    \textrm{s.t.} \quad
    (\ref{OP:NST-1}\textrm{a}):& \quad\left( \tau _{b}^{NS}-\tau _{p}^{NS} \right) P_{\Sigma}\le \tau _{p}^{NS} \varPhi _{\Sigma}\le E_{b,\max},\\
    (\ref{OP:NST-1}\textrm{b}):& \quad \tau _{p}^{NS}\leq\tau _{b}^{NS}\leq   \tau _{b,\max}^{NS}  ,\\
    (\ref{OP:NST-1}\textrm{c}):& \quad\lambda _{n,\min}\leqslant \lambda _n\leqslant \lambda _{n,\max},\\
    (\ref{OP:NST-1}\textrm{d}):& \quad0<\lambda _n\tau _{b}^{NS}<1,\\
    (\ref{OP:NST-1}\textrm{e}):& \quad\varUpsilon _{n}^{NS}\geq\varUpsilon _{min},
  \end{aligned}
\end{equation}
where the transmit power $\varPhi _{t,n}^{NS}$ and the EE of the system $\varUpsilon _{n}^{NS}$ are obtained by (\ref{OP:EN-1-ee}) and (\ref{OP:NST-1}).\par
According to Lemma \ref{lemma:4}, we obtain the optimal solutions of $\mathbf{P}^{NM}$ and $\mathbf{P}^{NS}$ via Algorithm \ref{alg:ELS}. Furthermore, in order to mitigate the computational complexity of devices, Algorithm \ref{alg:CCP} is used for solving $\mathbf{P}^{NM}$ and $\mathbf{P}^{NS}$, respectively. 
\section{Simulation Results And Discussions}
\label{sec:Simulation Results And Discussions}
This section presents numerical results for verifying the performance of both sleep-scheduling policies. We also use the three multiple access protocols without sleep-scheduling as our benchmark designs.
\subsection{Parameters Settings}
{\color{black}{As for the $n$th MTCD, in our simulation scenario, we assume that the distance between each MTCD and the BS or power station is the same\footnote{We assume that all MTCDs are on the perpendicular bisector of the line segment between the power station and the BS.}, i.e., $d_{n}^{P-M}=d_{n}^{M-B}$. The distance set is defined as $\boldsymbol{d}^{M-B}=\left( d_{1}^{M-B},\cdots ,d_{N}^{M-B} \right) $, which represents a linearly spaced vector in the range of $[3,5]$m and satisfies $d_{1}^{M-B}<\cdots<d_{N}^{M-B}$.}}
The available bandwidth is 5MHz with a carrier center frequency of 470MHz, while the TGn path-loss factor of 2 is used\cite{bouzinis2020pareto}. The noise power spectral density is $N_0=-60$dBm/Hz. The efficiency of EH is $\eta_p=0.9$. The maximum transmission time is $\tau_{b,max}=40$ms, while the EH time is a constant 10ms. Furthermore, $\varPhi _{r,max}$ and $\varPhi_{t,max}$ are 4W and 400mW, respectively. The probability of an MTCD becoming active in a EH or transmission mode is $\rho=\lambda_n\tau_b$\cite{tian2006vacation}. {\color{black}{Therefore, we define the average power consumption for the protocols based on sleep-scheduling as follows:
\begin{equation}\label{eq:a-p-c}
  \begin{aligned}
    \overline{\varPhi}=\rho\left(\frac{\tau_b-\tau_p}{\tau_b}\varPhi _{t}+\varPhi _{w}\right)+\left(\frac{1-\rho}{\tau_s}\right)\left(\varPhi_{s}\tau_i+\varPhi_{sc}\tau_{sc}\right),
\end{aligned}
\end{equation}
where $\varPhi_{w}=100$mW, $\varPhi_{s}=10$mW and $\varPhi _{sc}=100$mW represent the power consumption in the active mode, idle mode and the switching cost, respectively. Moreover, we suppose $\tau_s=\tau_i+\tau_{sc}$, where $\tau_i=9\tau_{sc}$.}} Substituting $\varPhi_{s}=\varPhi_{w}$ into (\ref{eq:a-p-c}), we obtain the average power consumption of our benchmark designs. 

\begin{figure}[t]
  \centering 
  \subfigure[MV policy]{
  \includegraphics[width=3in]{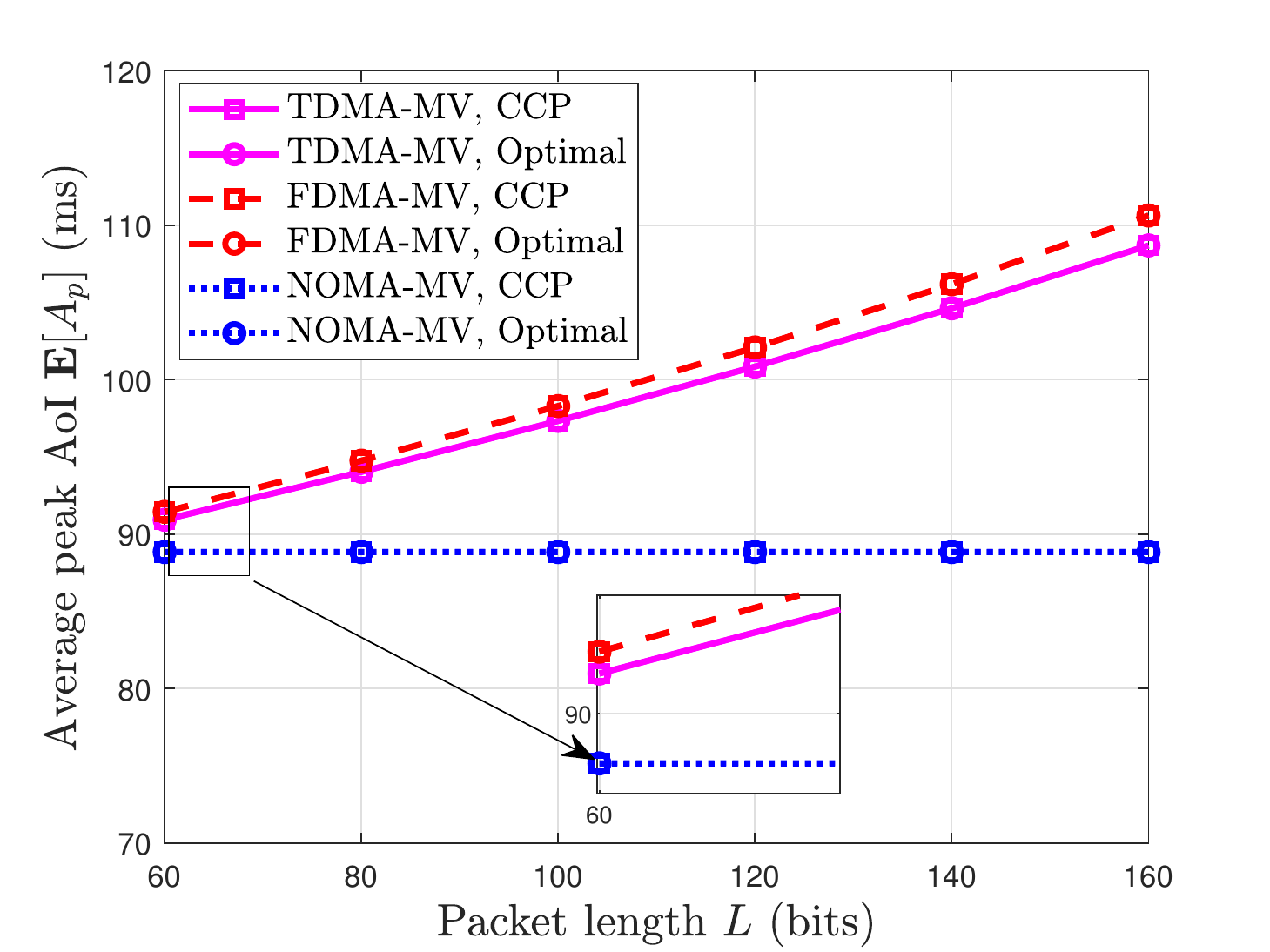}
  }\\
  \subfigure[ST policy]{
  \includegraphics[width=3in]{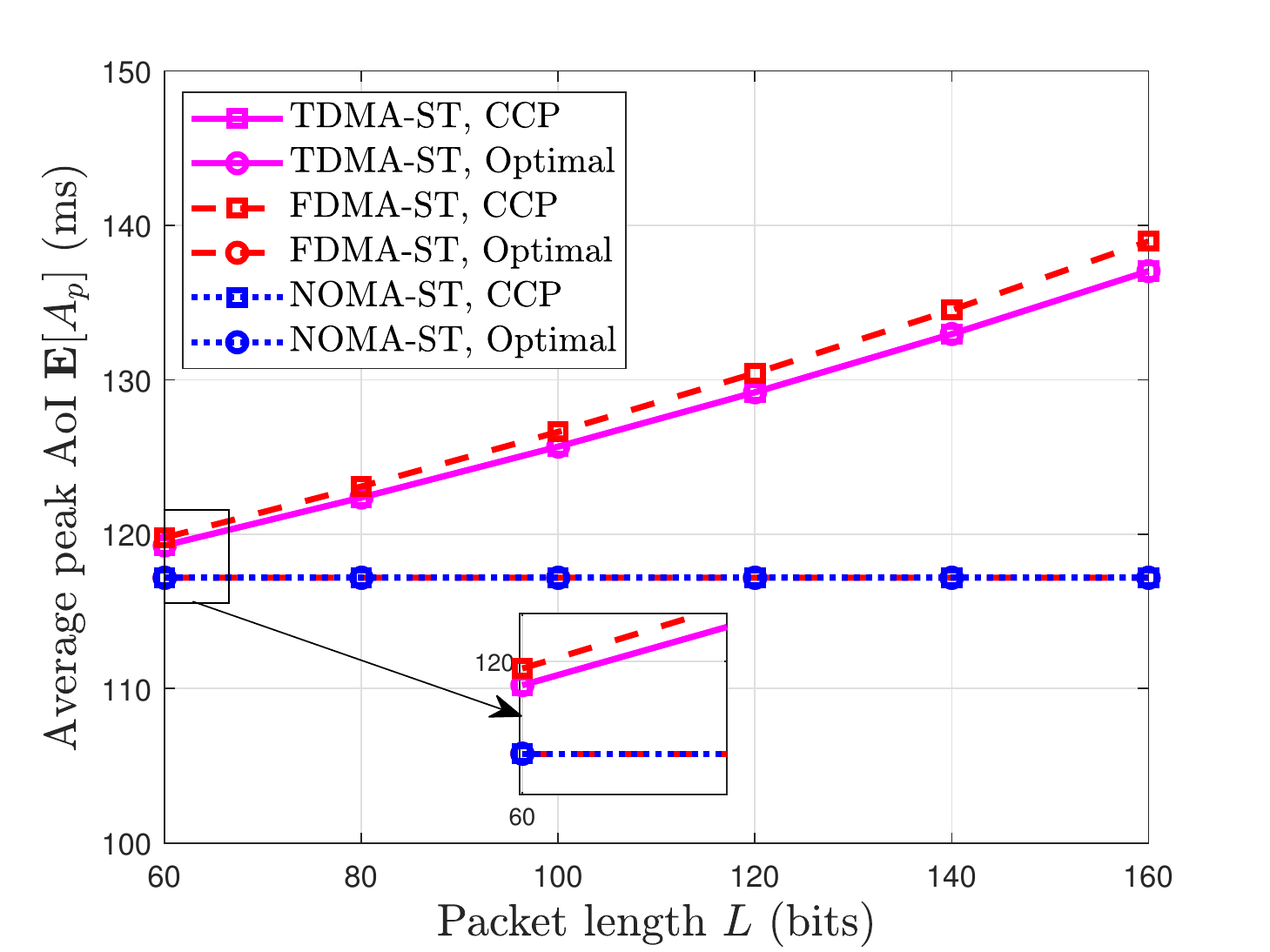}
  }%
  \caption{The average peak AoI versus $L$ under a pair of sleep-scheduling policies.}
  \label{fig:AoI_L_MV_ST}
\end{figure}

\begin{figure}[t]
  \centering
  \includegraphics[width=0.47\textwidth]{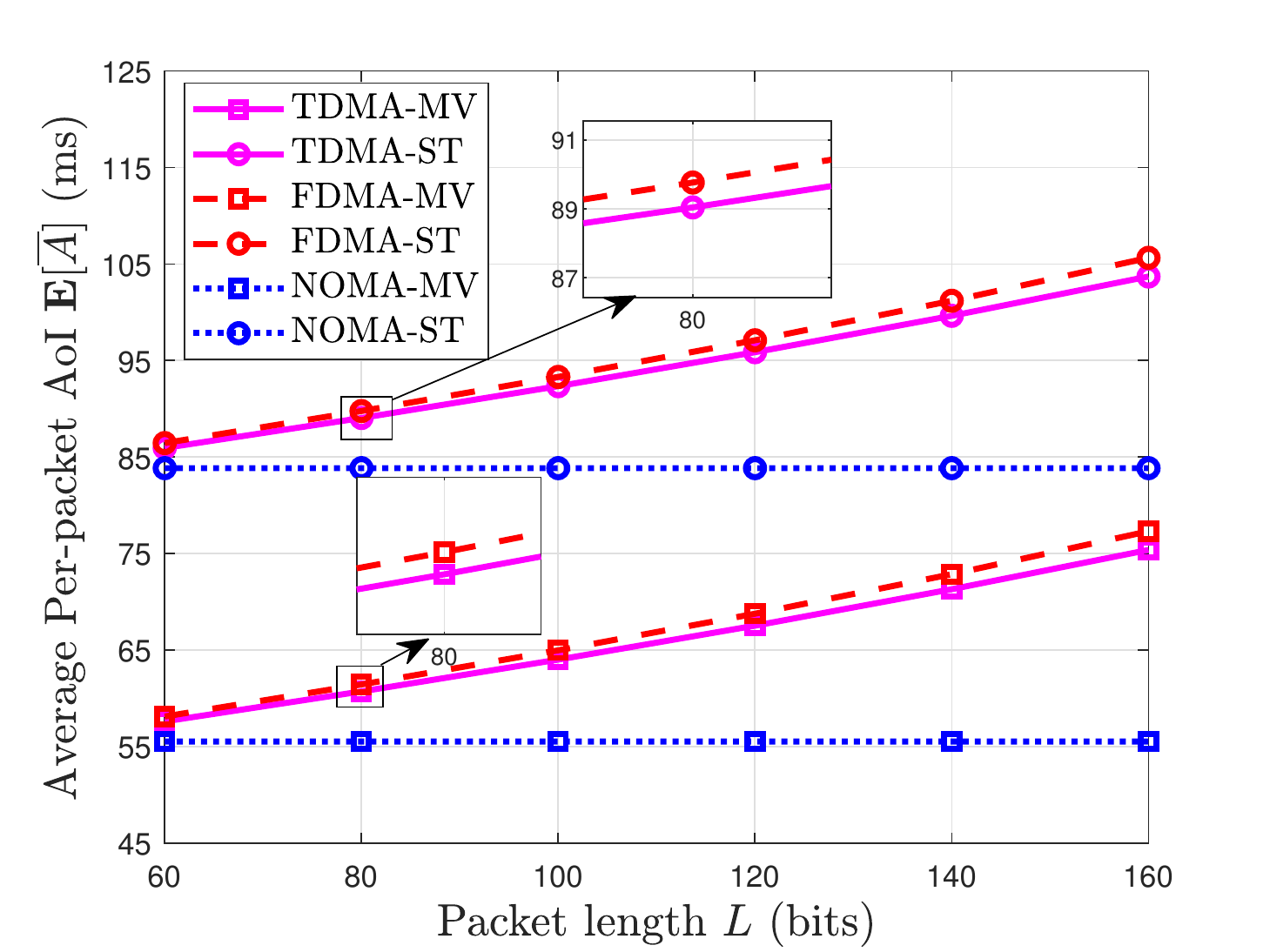}
  \caption{The average per-packet AoI versus $L$ under a pair of sleep-scheduling policies.}
  \label{fig:pp_AoI_L}
  \vspace{-6mm}
\end{figure}
\begin{figure}[h]
  \centering 
  \subfigure[MV policy]{
  \includegraphics[width=3in]{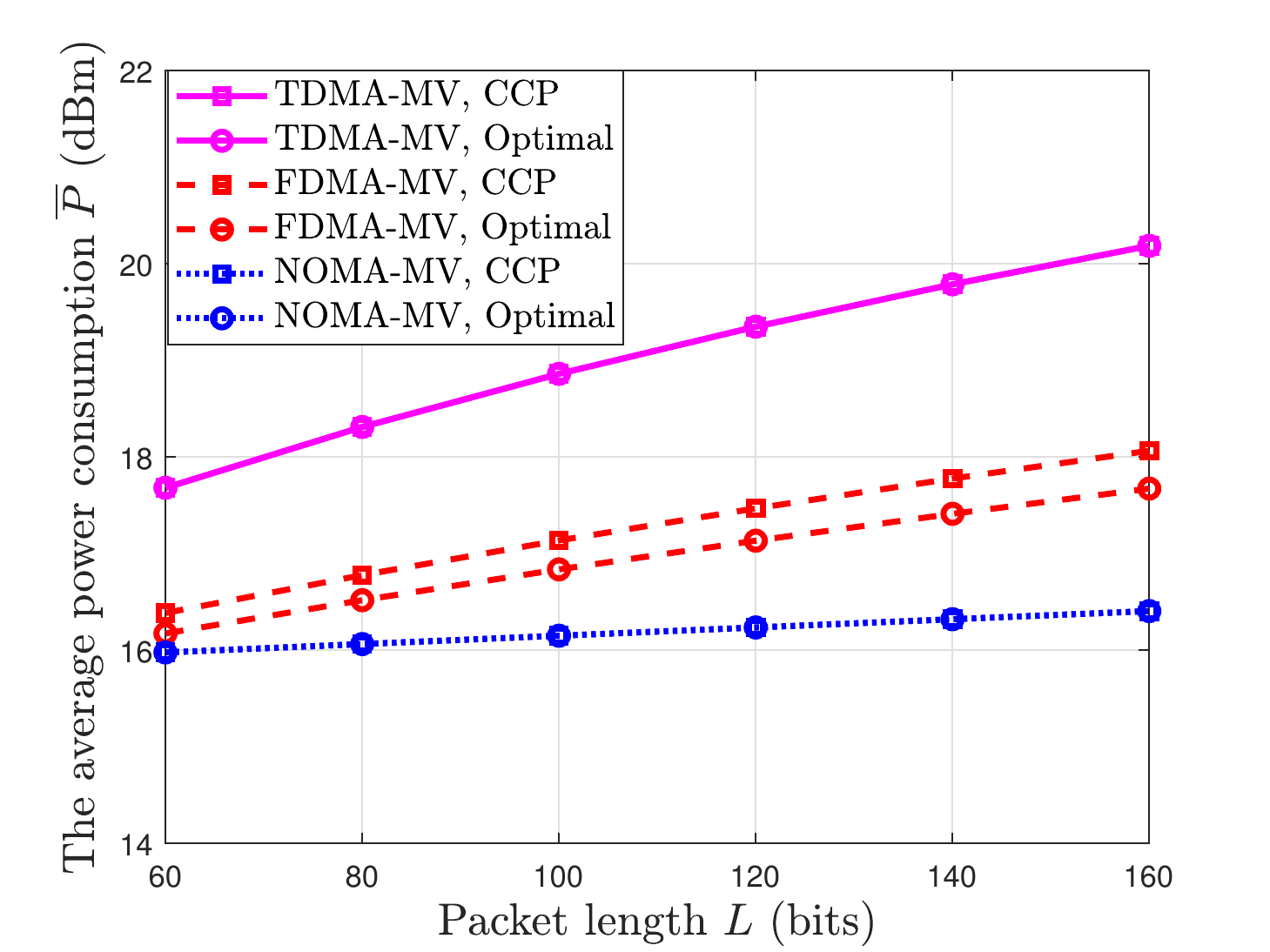}}\\%
  \subfigure[ST policy]{
  \includegraphics[width=3in]{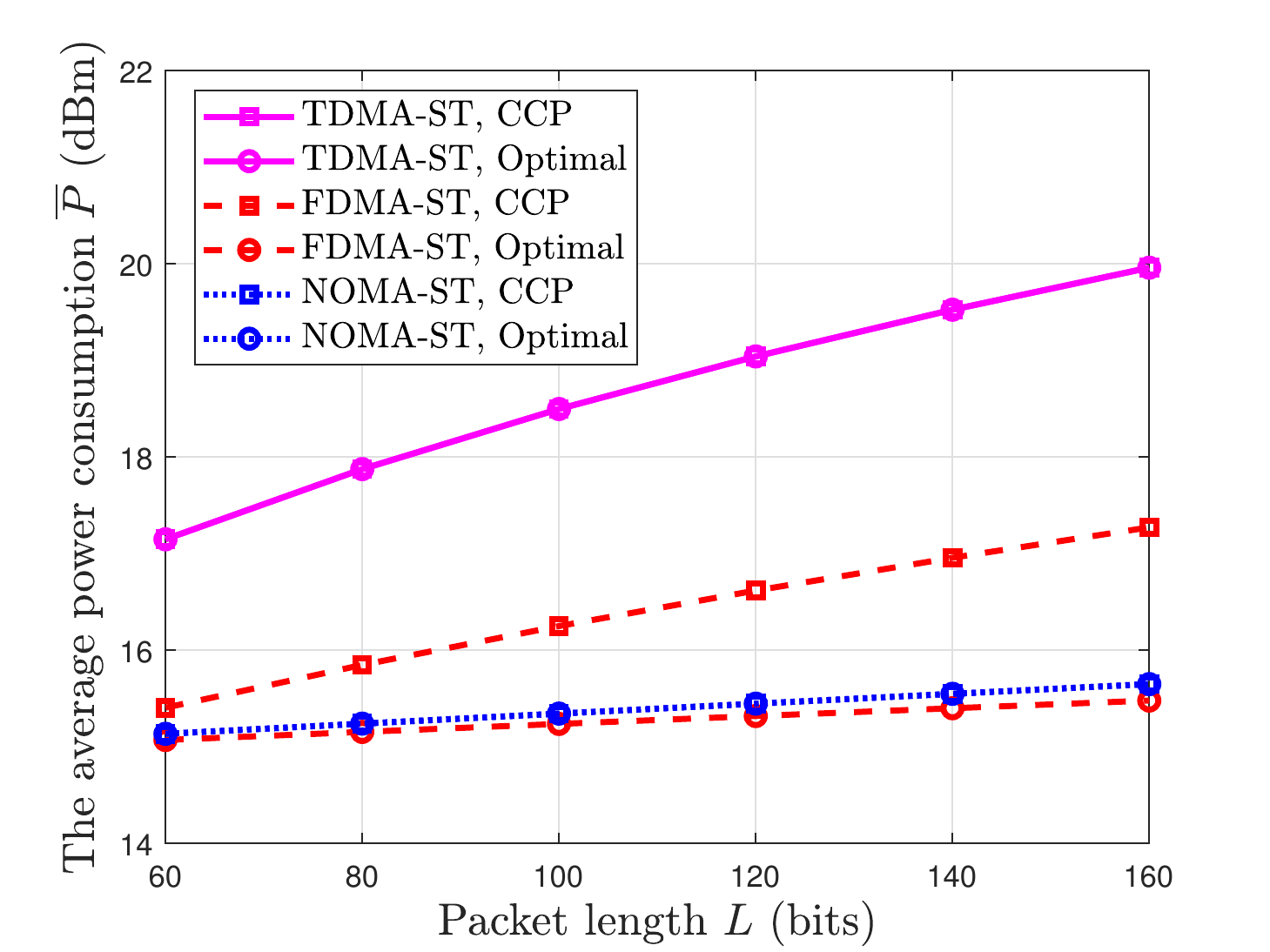}}%
  \caption{The average power consumption versus $L$ under a pair of sleep-scheduling policies.}
  \label{fig:EE_L_MV_ST}
\end{figure}

In Fig. \ref{fig:AoI_L_MV_ST}(a), we consider the scenario of 10 MTCDs relying on the MV policy, where the maximum status update rate is $\lambda_{max}=15$ packets per second. It is observed that for the different protocols relying on the increment of $L$, the average peak AoI increases under the MV policy. A large $L$ results in an increased throughput and tell-traffic load, hence leading to a higher peak AoI. Moreover, NOMA has a lower peak AoI than FDMA and TDMA, but the peak AoI of NOMA increases marginally when the gap between the channel capacity and the current throughput is substantial. Although CCP is a heuristic algorithm, it can be seen that the results based on CCP are very close to the optimal solutions. Fig. \ref{fig:AoI_L_MV_ST}(b) characterizes the average peak AoI using the ST policy. Similar to the protocols based on the MV policy, it can be observed that the freshness of data is directly related to the packet length and that NOMA under the ST policy is superior to both TDMA and FDMA in terms of its peak AoI. Observe that harnessing the CCP method in the NOMA protocol strikes a trade-off between the average peak AoI attained and the computational complexity imposed. {\color{black}{Additionally, Fig. \ref{fig:pp_AoI_L} illustrates the results of average per-packet AoI. According to Eq. (\ref{eq:per-packet AoI}), the relationship between the average per-packet AoI and the average peak is linear in terms of the same status update rate $\lambda$, and thus Fig. \ref{fig:pp_AoI_L} follows the same trend as Fig. \ref{fig:AoI_L_MV_ST}. Furthermore, it can be observed that when $L$ is less than $160$ bits, TDMA-MV and FDMA-MV have a better per-packet AoI performance than NOMA-ST, because the delay of NOMA is degraded by the ST policy.}}

In Fig. \ref{fig:EE_L_MV_ST}(a), we evaluate the average energy consumption $\overline{P}$ versus the packet length. Observe that a shorter packet is beneficial in terms of saving energy. Moreover, relying on our proposed MV policy, NOMA outperforms the other multiple access protocols. However, upon using the CCP method, FDMA using the MV policy exhibits a reduced power for $60<L<160$ bits. Fig. \ref{fig:EE_L_MV_ST}(b) shows the impact of the packet length on the average power consumption $\overline{P}$. With the increment of $L$, more status updates have to be transmitted by the MTCDs, which requires more energy from the power station to reduce the peak AoI. For reducing the computational complexity, the following experiments are conducted relying on CCP method rather than the exact linear search.
\par
\subsection{Performance Versus Maximum Status Update Rate}
\begin{figure*}
  \centering
  \subfigure[TDMA protocol]{
  \includegraphics[width=5.0cm]{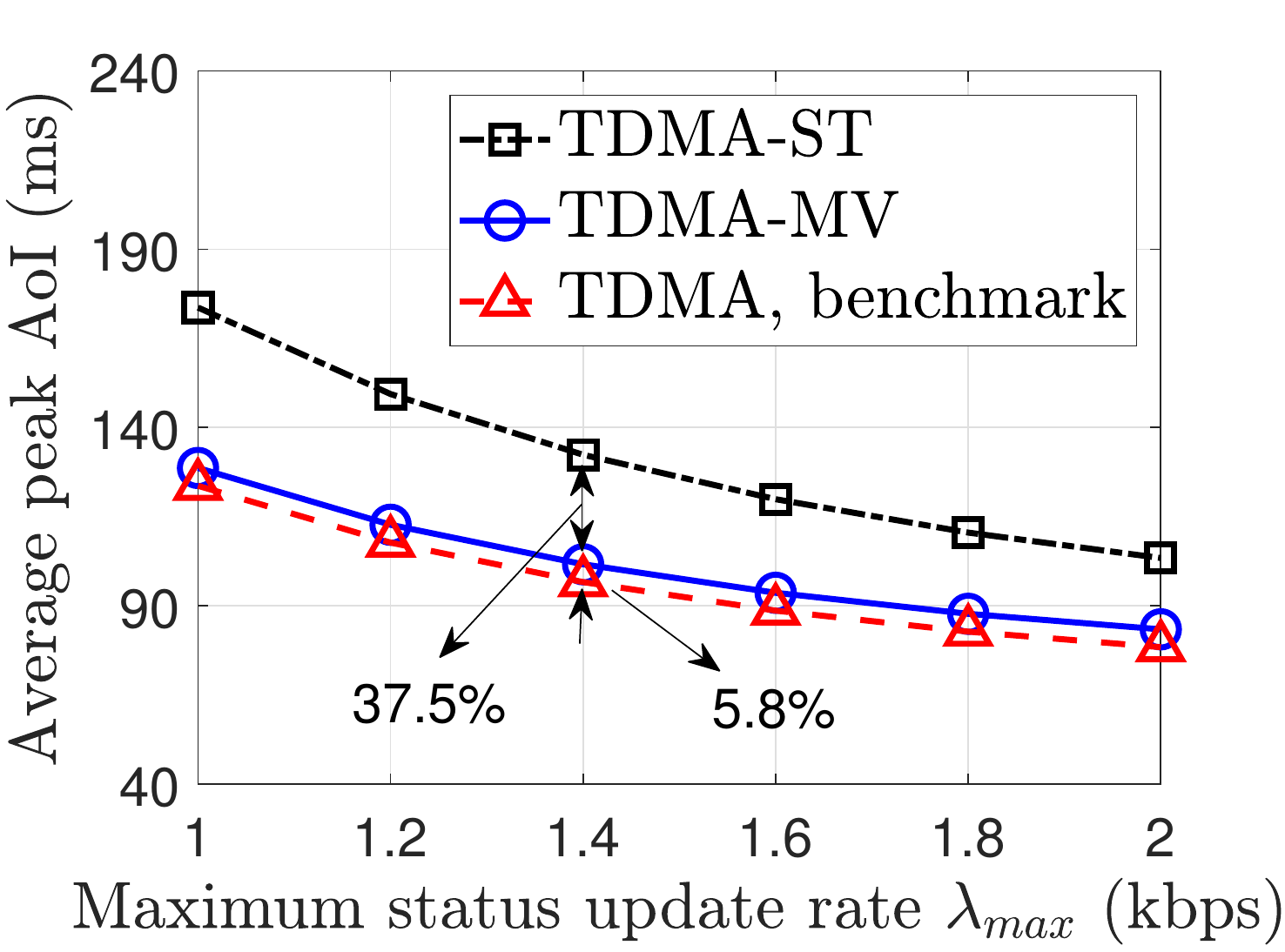}
  }
  \subfigure[FDMA protocol]{
  \includegraphics[width=5.0cm]{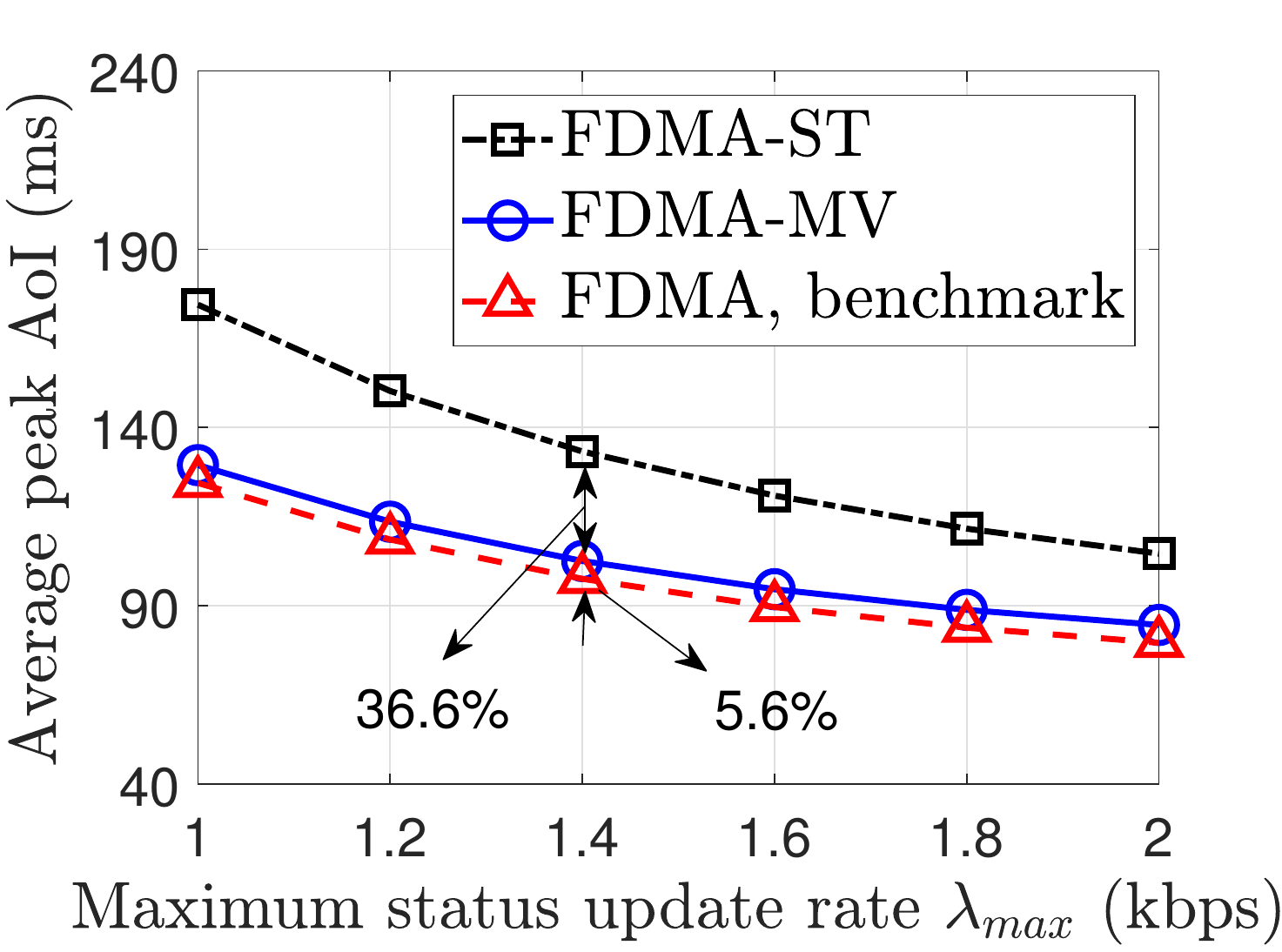}
  }
  \subfigure[NOMA protocol]{
  \includegraphics[width=5.0cm]{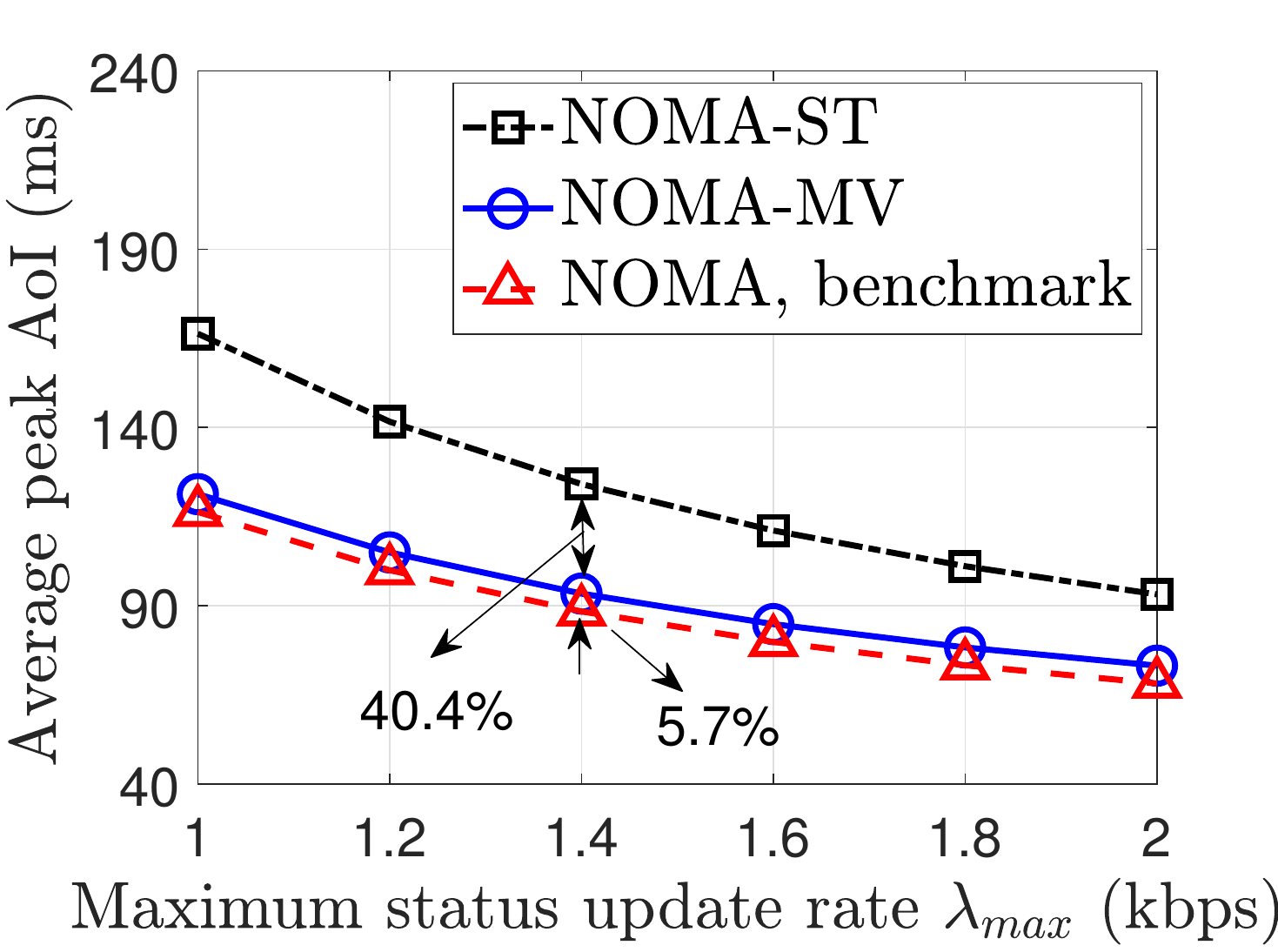}
  }
  \caption{The average peak AoI versus different status update rate constraints.}
  \label{fig:AoI-TC}
\end{figure*}
\begin{figure*}
  \centering
  \subfigure[TDMA protocol]{
  \includegraphics[width=5.0cm]{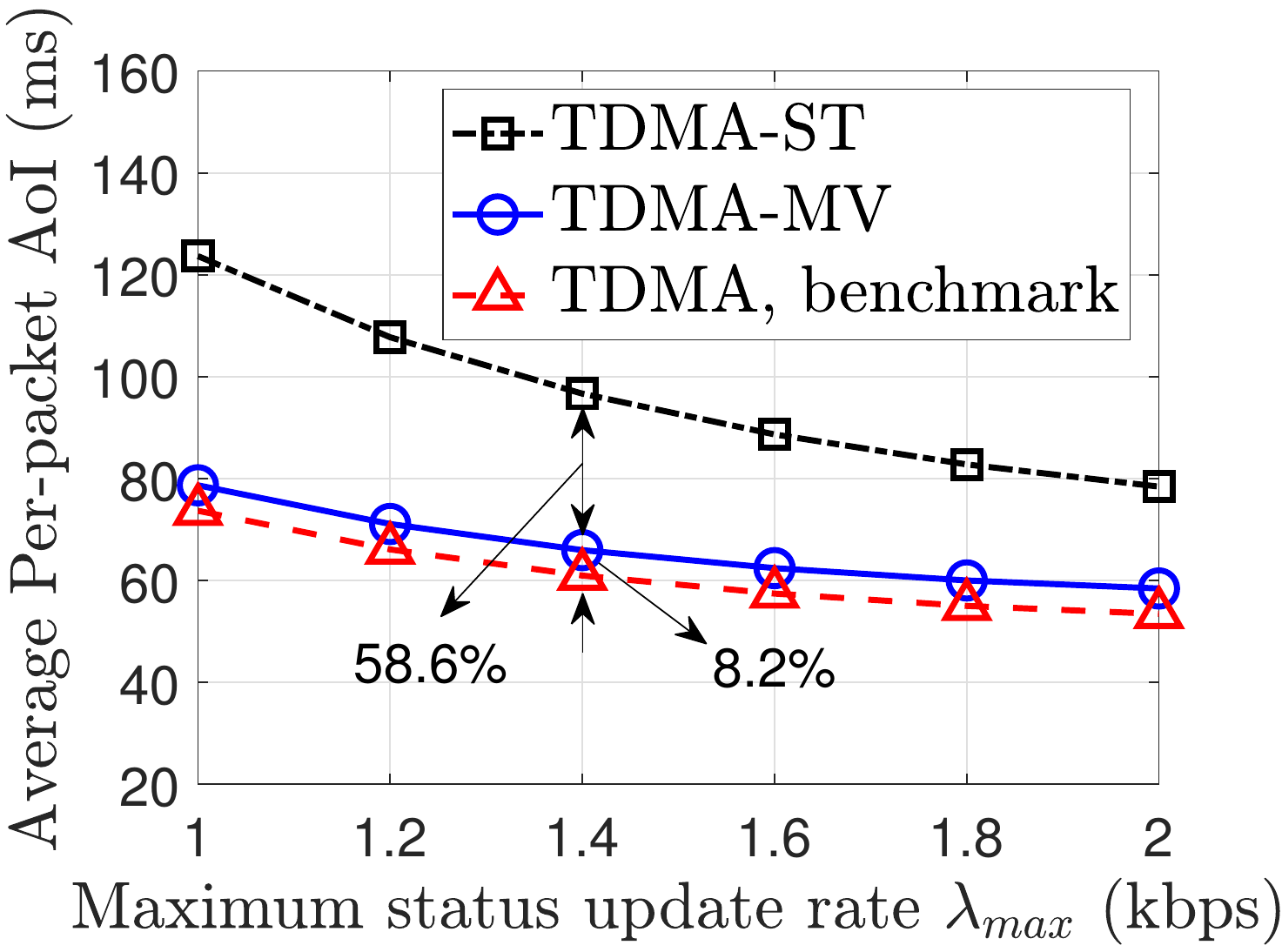}
  }
  \subfigure[FDMA protocol]{
  \includegraphics[width=5.0cm]{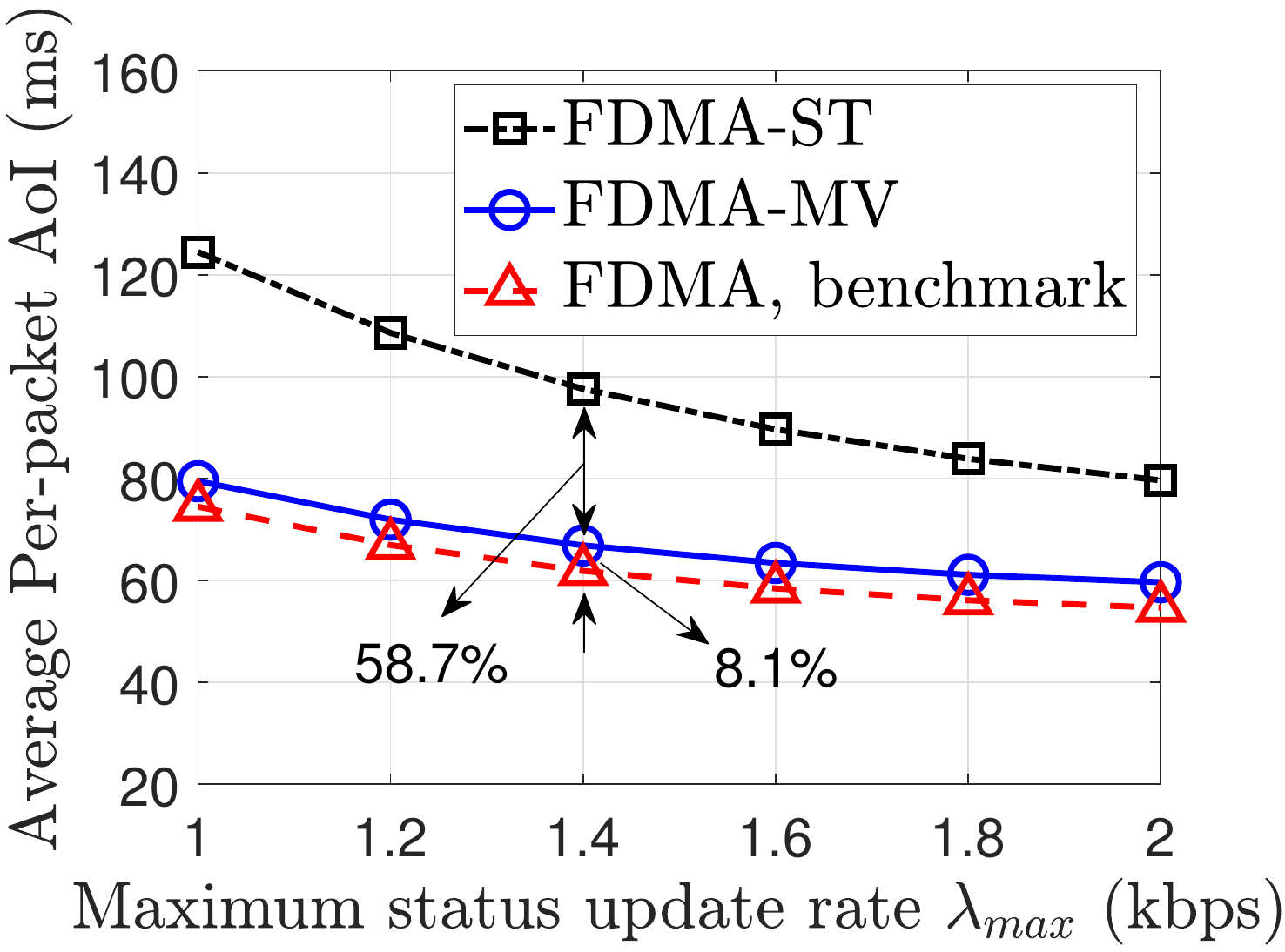}
  }
  \subfigure[NOMA protocol]{
  \includegraphics[width=5.0cm]{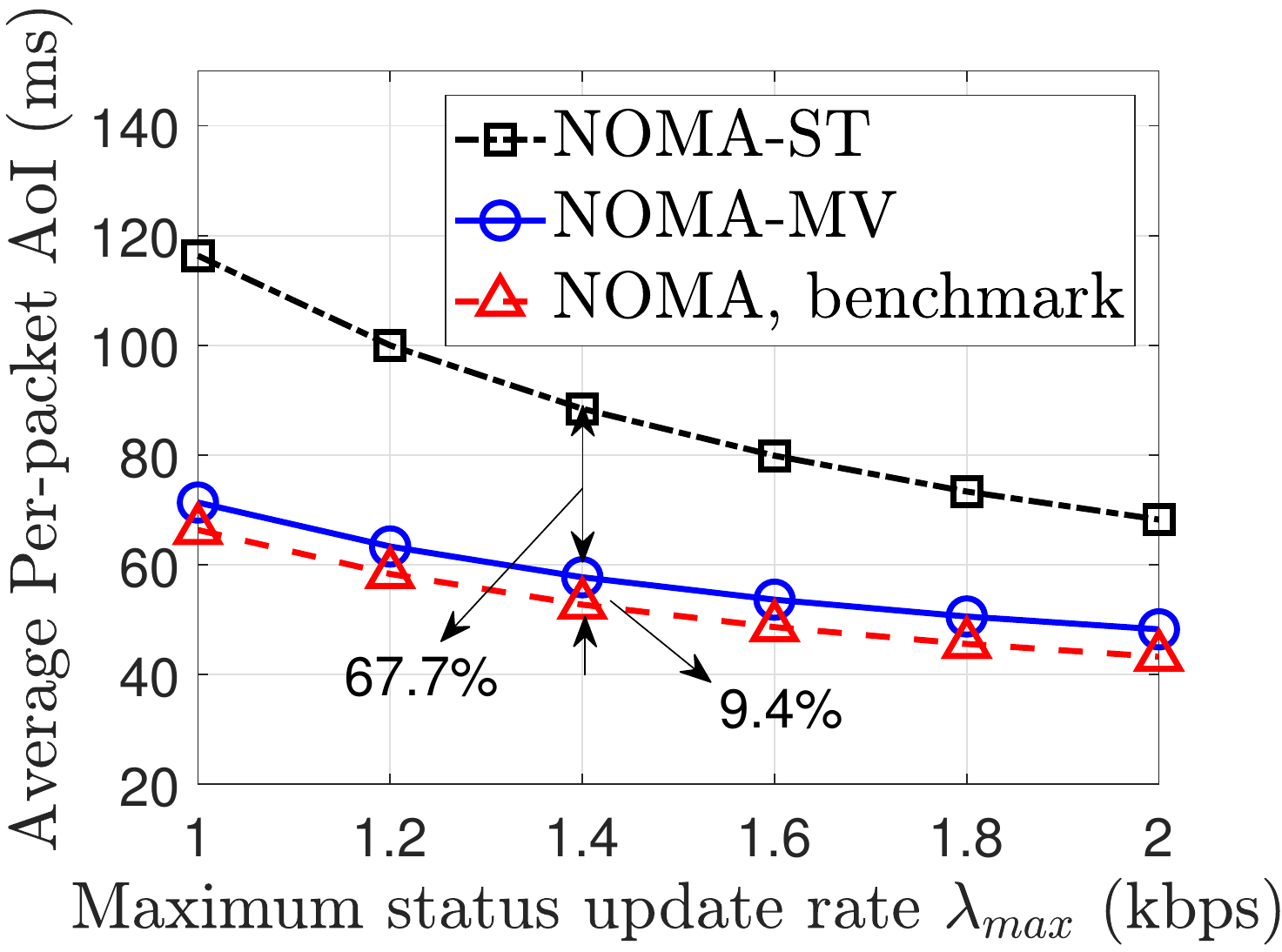}
  }
  \caption{The average per-packet AoI versus different status update rate constraints.}
  \label{fig:AoI-TTC}
\end{figure*}
\begin{figure*}[h]
  \centering
  \subfigure[TDMA protocol]{
  \includegraphics[width=5.0cm]{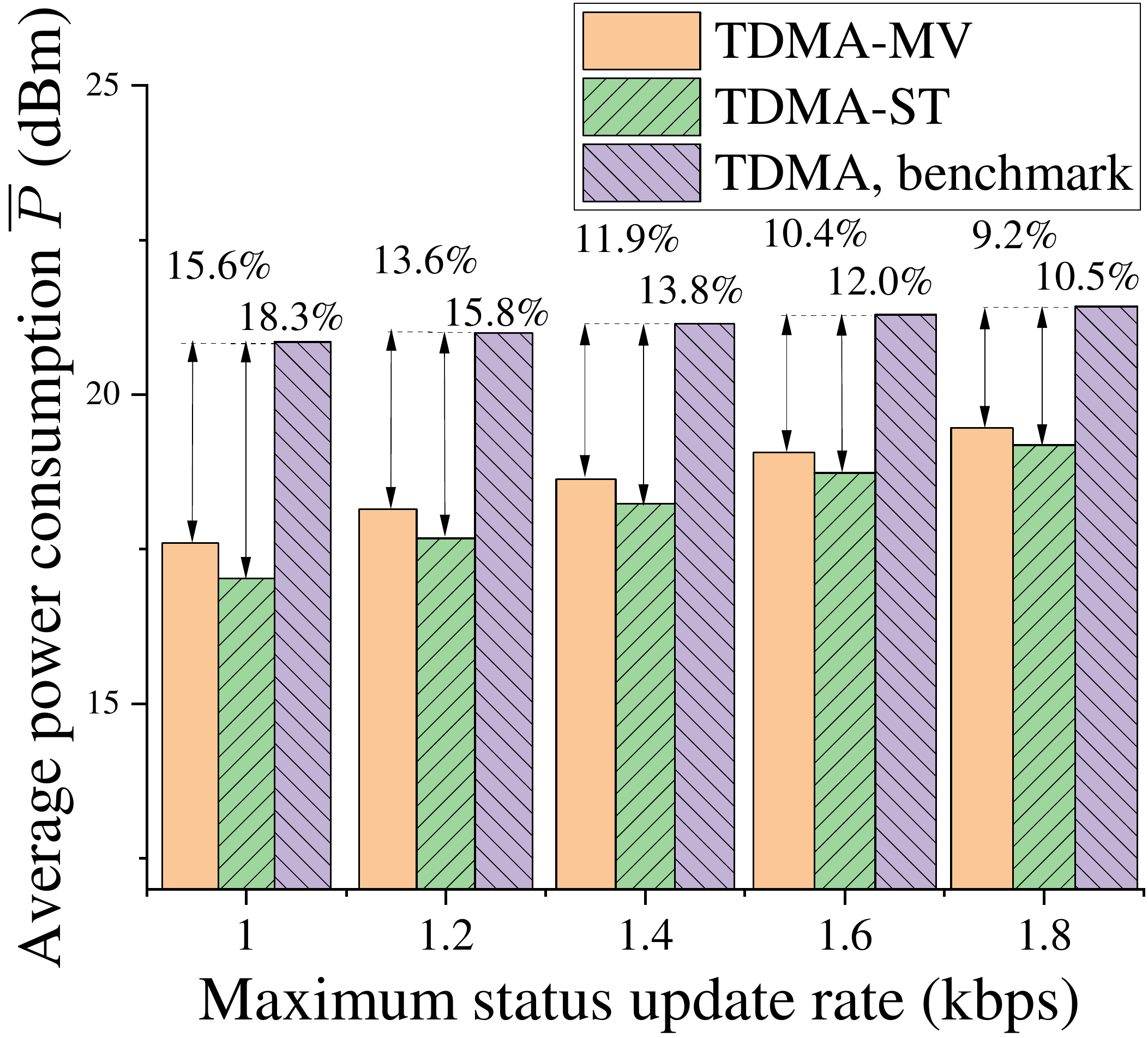}\label{fig:EE-TC-1}
  }
  \subfigure[FDMA protocol]{
  \includegraphics[width=5.0cm]{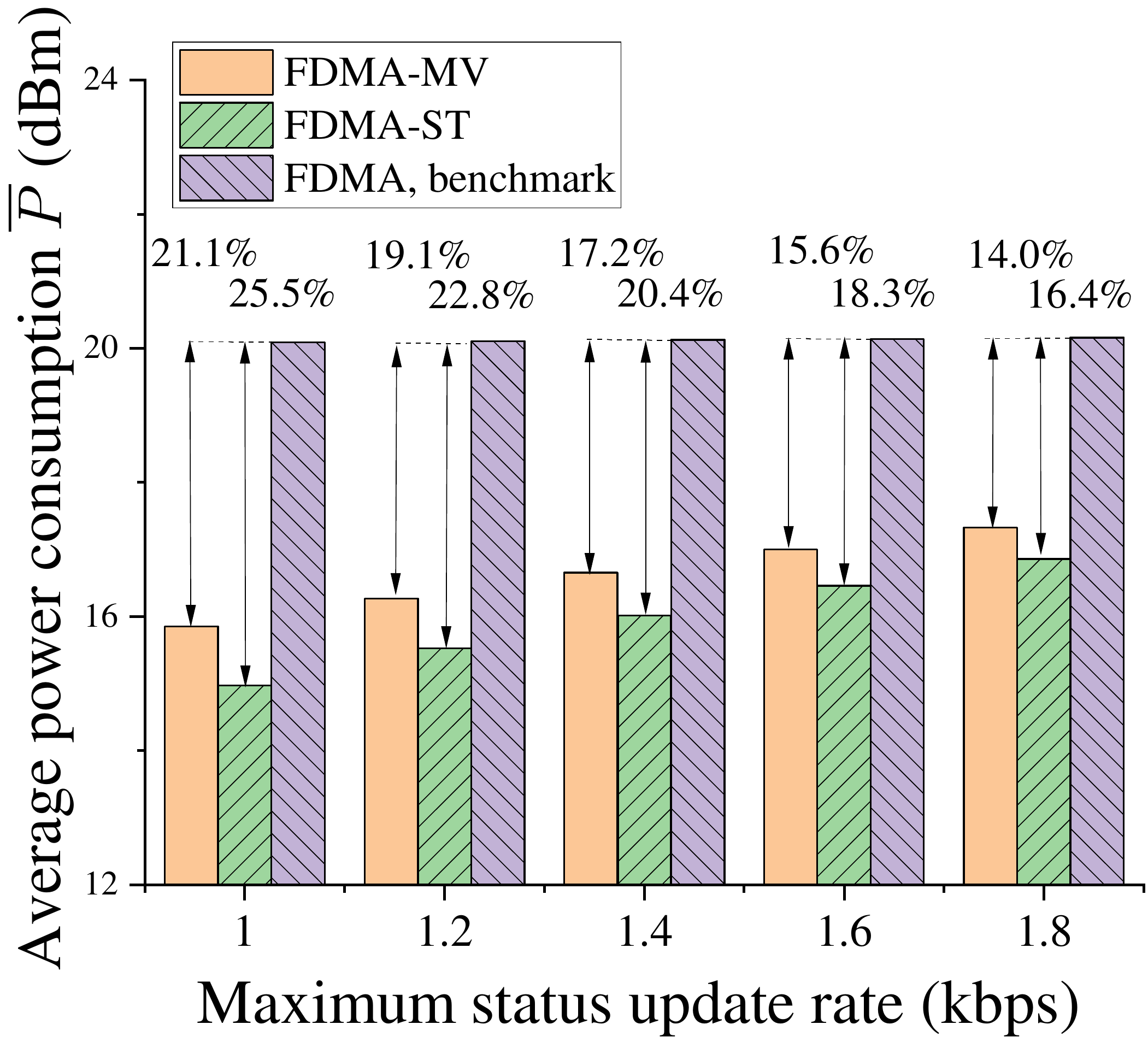}\label{fig:EE-TC-2}
  }
  \subfigure[NOMA protocol]{
  \includegraphics[width=5.0cm]{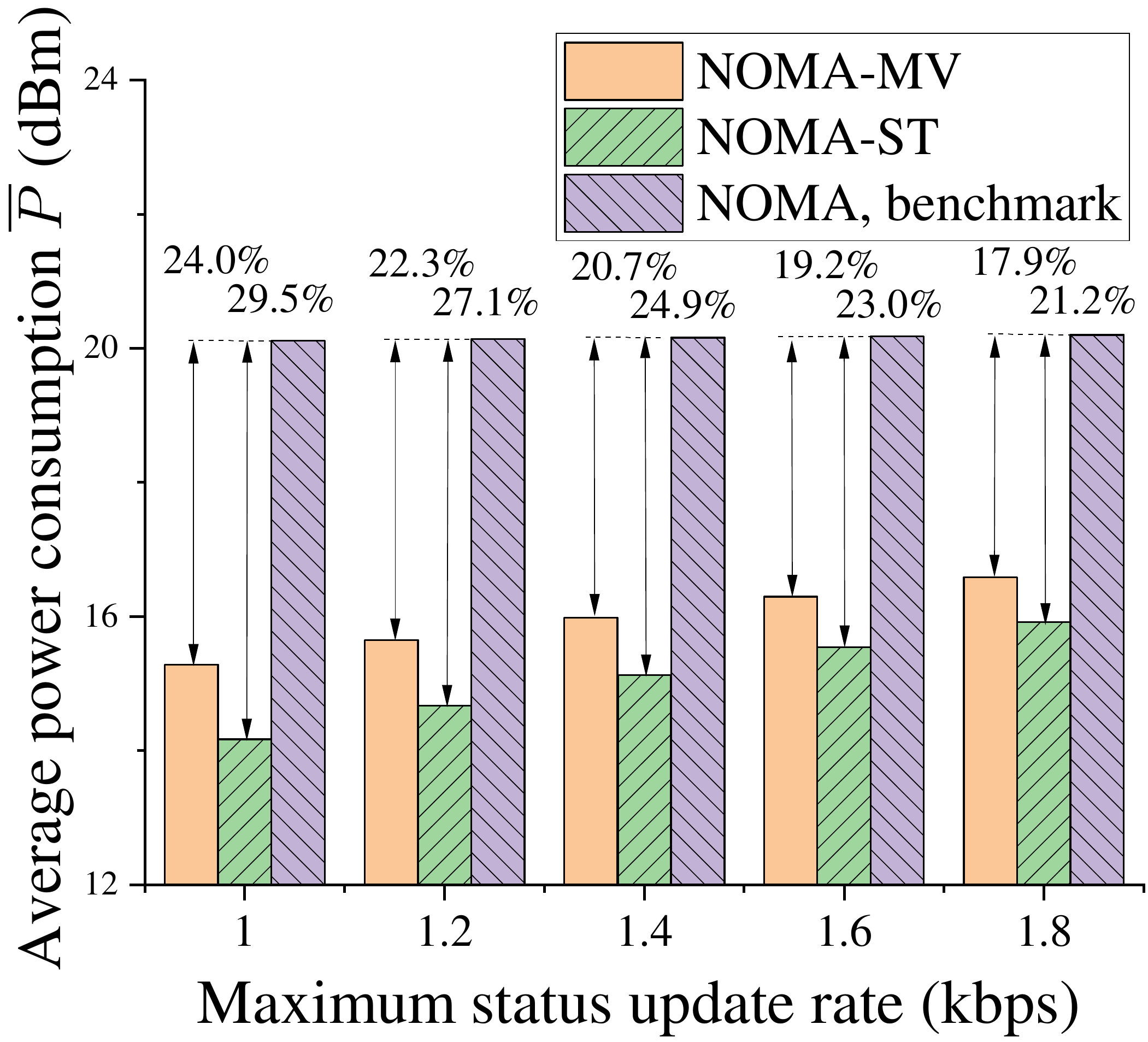}\label{fig:EE-TC-3}
  }
  \caption{The average power consumption versus different status update rate constraints.}
  \label{fig:EE-MA}
\end{figure*}
{\color{black}{This subsection studies the impact of the maximum status update rate of both policies on the AoI and on the energy consumption, where $N=10$ and $L=100$bits, for all three multiple access protocols. Fig. \ref{fig:AoI-TC}(a) and Fig. \ref{fig:AoI-TC}(b) indicate that the networks based on TDMA-MV and FDMA-MV have a lower peak AoI, when the maximum status update rate increases. This is because the status updates become infrequent when the maximum status update rate $\lambda_{max}$ is low, which in turn increases the average peak AoI. Since NOMA has a better energy efficiency than the other protocols, its average peak AoI is monotonically reduced in Fig. \ref{fig:AoI-TC}(c). Moreover, Fig. \ref{fig:AoI-TC} also demonstrates that the sleep-scheduling policies degrade the AoI for the sake of reducing the energy consumption. At $\lambda_{max}=1.4$ kbps, the ST policy imposes 37.5\%, 36.6\% and 40.4\% average peak AoI degradation in TDMA, FDMA and NOMA, respectively. By contrast, the average peak AoI values of the MV policy only increase by 5.8\%, 5.6\% and 5.7\% relying on TDMA, FDMA and NOMA, respectively. Therefore, the MV policy is more suitable for a low-delay scenarios than the ST policy. Additionally, Fig. \ref{fig:AoI-TTC} illustrates the average per-packet AoI vs. the status update rate $\lambda_{max}$, which has a similar AoI trend as Fig. \ref{fig:AoI-TC}. Specifically, at $\lambda_{max}=1.4$ kbps, the average per-packet AoI increases by 58.6\%, 58.7\% and 67.7\% in TDMA, FDMA and NOMA, respectively. As for the MV policy, the AoI degradations are 8.2\%, 8.1\% and 9.4\% for TDMA, FDMA and NOMA, respectively. Observe that the sleep-scheduling policies have a more grave impact on the per-packet AoI than on the peak AoI. Furthermore, the AoI degradation of NOMA is higher than that of the other protocols.
\par
Fig. \ref{fig:EE-MA} demonstrates the performance of the MTCD's average power consumption for the different policies. As shown in the figure, compared to the benchmark protocols, the power consumption is reduced by both the ST and MV policies. Moreover, the average power consumption increases with $\lambda_{max}$, since the packets within the queue are transmitted more frequently to maintain the queue's stability, which leads to an increased energy consumption. Specifically, when $\lambda_{max}=1.4$ kbps and the MV policy is adopted, the average energy consumption of MTCDs can be reduced by 11.9.5\%, 17.2\% and 20.7\% for TDMA, FDMA and NOMA, respectively. As for the ST policy, the average energy consumption can be reduced by 13.8\%, 20.4\% and 24.69\% for TDMA, FDMA and NOMA, respectively. Therefore, it can be concluded that the sleep-scheduling policies reduce the energy consumption at the cost of an increased AoI. Although the ST policy is superior to the MV policy in terms of its energy dissipation, the MV policy may be better suited for near real-time status update scenario.}}
\subsection{Performance Versus the Number of MTCDs}
\begin{figure}[t]
  \centering 
  \subfigure[Average peak AoI]{
  \begin{minipage}[t]{0.5\linewidth}
    \centering 
  \includegraphics[width=1.85in]{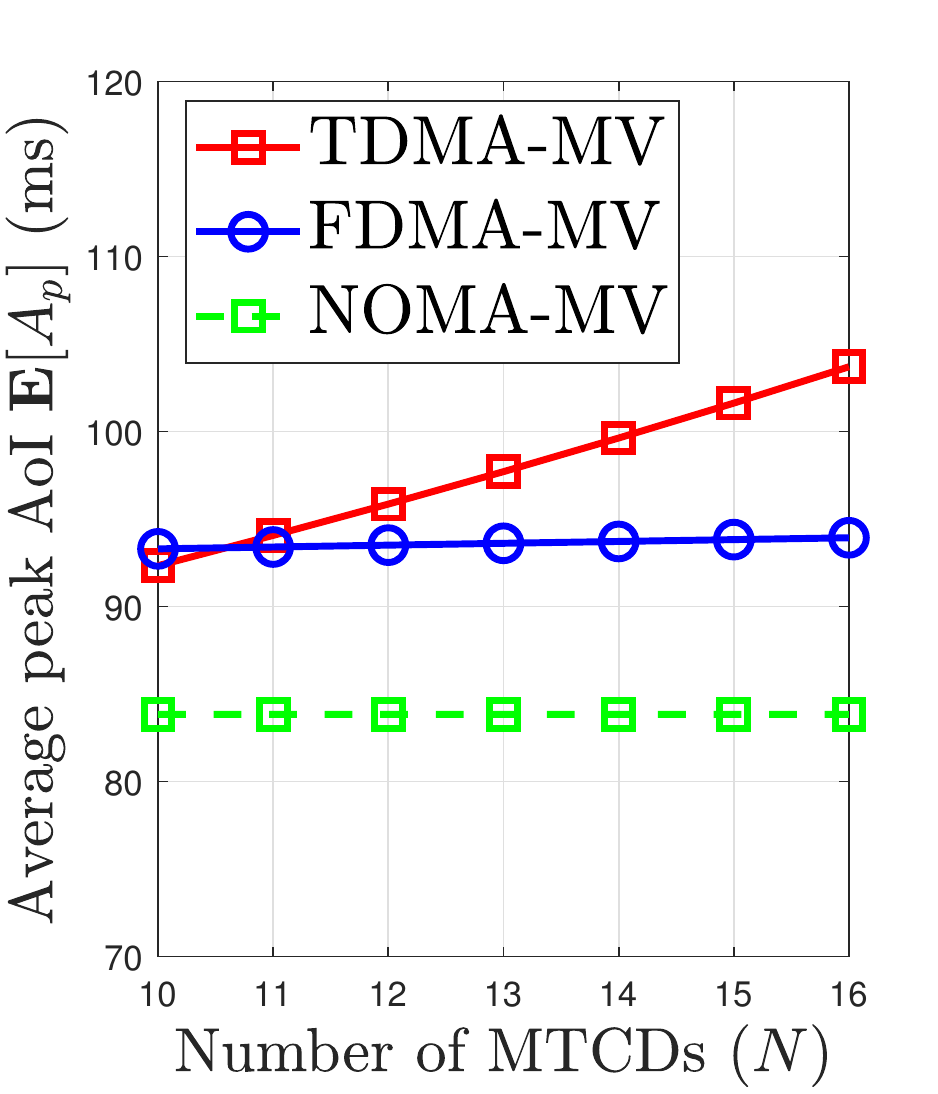}
  \end{minipage}%
  }%
  \subfigure[Average power consumption]{
  \begin{minipage}[t]{0.50\linewidth}
    \centering 
  \includegraphics[width=1.85in]{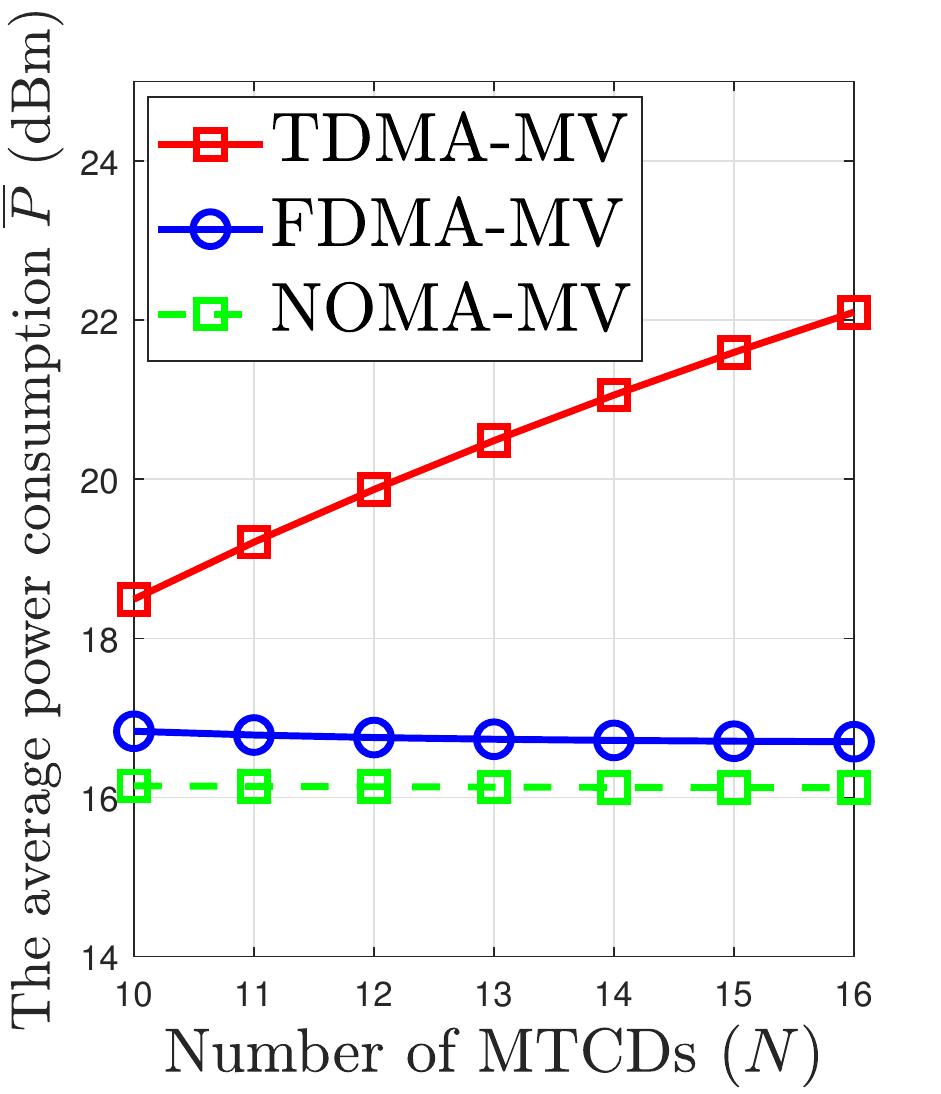}
  \end{minipage}%
  }%
  \flushleft 
  \caption{The peak AoI and the sum of average power consumption versus $N$ under MV policy.}
  \label{fig:AoI-EE-MV}
  \vspace{-6mm}
\end{figure}
{\color{black}{In Fig. \ref{fig:AoI-EE-MV}, we evaluate the average peak AoI and the sum of the average power consumption (i.e. $P_{\Sigma}=N\cdot\overline{P}$) for different number of MTCDs $N$, where we have $L=100$bits and $\lambda_{max}=15$. Observe that upon increasing $N$, the optimal average peak AoI increases. The reason for this trend is that the total throughput of the network rises sharply when the number of MTCDs is increased, hence the packets are more likely to be queued for a long time and require more energy from the power station. Furthermore, for the MV policy, the average peak AoI of FDMA is better than that of TDMA, and NOMA always has the edge. That is because the spectral resources of FDMA and NOMA are capable of supporting an increased number of MTCDs, but the TDMA has a limited number of time slots.}}
\section{Conclusions}
\label{sec:Conclusion}
The family of EH-aided NGMA systems relying on the multiple vacation policy and start-up threshold policy has been investigated and closed-form expressions for the peak AoI of both policies have been derived. Since the latency caused by sleep-scheduling leads to the degradation of AoI, we have formulated the associated time slot allocation problems for minimizing the average peak AoI and obtained the optimal solutions of the resultant non-convex problems by an exact linear search method. For the sake of reducing the computational complexity, we have formulated the original problems as DC problems, which can be solved by CCP based heuristic algorithms. Our simulation results have shown that the proposed sleep-scheduling policies are more energy-efficiency than the benchmark protocols. Additionally, under the premise of fair rate allocation among MTCDs, NOMA outperforms both TDMA and FDMA in terms of its peak AoI and power consumption.
\begin{appendices}
\section{Proof of Lemma 2}\label{proof:lemma-2}
\begin{Proof}
Let $\boldsymbol{\pi_{\varPi}}=\left(\pi_1,\pi_2,\cdots\right)$ be the stationary distribution of $\varPi_n^{ST}$, and $\boldsymbol{e}=(1,1,\cdots)^\mathrm{T}$. According to the equilibrium equation of $\boldsymbol{\pi_{\varPi}}\boldsymbol{P}_0=\boldsymbol{\pi_{\varPi}}$ and Eq. (\ref{eq:TDMA-ST-3}), we can formulate the stationary distribution as follows:
\begin{equation}\label{eq:lemma-2-1}
  \begin{aligned}
    \pi _m=\begin{cases}
      \sum_{i=1}^{m+1}{\pi _i\theta _{m+1-i}} &  0\leqslant k\leqslant M-2,\\
      \pi _0\theta _{m-M+1}+\sum_{i=1}^{m+1}{\pi _i\theta _{m+1-i}} &  k\geqslant M-1.\\
    \end{cases}
\end{aligned}
\end{equation}
Therefore, the p.g.f. of the stationary distribution $\boldsymbol{\pi_{\varPi}}(z)$ can be expressed as follows:
\begin{equation}\label{eq:lemma-2-2}
  \begin{aligned}
    \boldsymbol{\pi }_{\varPi}\left( z \right)&=\sum_{i=M-1}^{+\infty}{z^i\left[ \pi _0\theta _{i-M+1}+\sum_{j=1}^{i+1}{\pi _j\theta _{i+1-j}} \right]}+\sum_{i=0}^{M-2}{z^i\sum_{j=1}^{i+1}{\pi _j\theta _{i+1-j}}}\\
    &=\frac{\left[ \boldsymbol{\pi }_{\varPi}\left( z \right) -\pi _0 \right] +\pi _0z^M}{z}e^{-\lambda _n\tau _b\left( 1-z \right)}.
\end{aligned}
\end{equation}
Upon substituting $\boldsymbol{\pi_{\varPi}}(1)=1$ into the above equation, we may arise at $\pi_0=\left(1-\lambda_n \tau_b \right)M^{-1}$. Therefore, Eq. (\ref{eq:lemma-2-2}) can be simplified as
\begin{equation}\label{eq:lemma-2-3}
  \begin{aligned}
    \boldsymbol{\pi }_{\varPi}\left( z \right)=\frac{\left( 1-\lambda _n\tau _b \right) \left( 1-z \right) \left( 1-z^M \right) e^{-\lambda _n\tau _b\left( 1-z \right)}}{M\left[ e^{-\lambda _n\tau _b\left( 1-z \right)}-z \right] \left( 1-z \right)}.
\end{aligned}
\end{equation}
Furthermore, the average number of the packets generated in the $n$th MTCD can be expressed as
\begin{equation}\label{eq:lemma-2-4}
  \begin{aligned}
\mathbb{E} \left[ \varPi_n^{ST} \right] &=\left.\dfrac{d\boldsymbol{\pi }_{\varPi}\left( z \right)}{dz}\right|_{z=1}=\lambda _n\tau _b+\frac{\left( \lambda _n\tau _b \right) ^2}{2\left( 1-\lambda _n\tau _b \right)}+\frac{M-1}{2}.
\end{aligned}
\end{equation}
Based on Little's Law, we can express the average system delay of the packets in the $n$th MTCD as $\mathbb{E} \left[ D_n^{ST} \right]=\lambda_n^{-1}\mathbb{E} \left[ \varPi_n^{ST} \right]$.
\end{Proof}
\section{Proof of Lemma 4}\label{proof:lemma-3}
\begin{Proof}
  The variables $\lambda_n, \tau _b^{TM},\tau _s^{TM}$ and ${A}_p ^{MV}\left( \lambda_n ,\tau _b^{TM},\tau _s^{TM} \right)$ are defined in Section \ref{sec:PAoI_TDMA-IB}. Moreover, we decompose the original objective function into the pair of functions $f_{1}^{TM}$ and $f_{2}^{TM}$, namely ${A}_p ^{MV}=f_{1}^{TM}+f_{2}^{TM}$, which are given by
  \begin{equation}\label{eq:proof-3-1}
    \begin{aligned}
      \begin{cases}
	f_{1}^{TM}\left( \boldsymbol{x}^{TM} \right) =\frac{\tau _{s}^{TM}}{2}+\frac{\tau _{b}^{TM}}{2}+\frac{1}{\lambda _n},\\
	f_{2}^{TM}\left( \boldsymbol{x}^{TM} \right) =\frac{\tau _{b}^{TM}}{2\left( 1-\lambda _n\tau _{b}^{TM} \right)}.
  \end{cases} 
  \end{aligned}
  \end{equation}
While $\frac{1}{\lambda _n}$ denotes the power function of $\lambda_n$ when $\lambda_n>0$. Therefore, $f_{1}^{TM}$ is convex on $\mathbf{R}_{+}^{2}$, because the nonnegative, nonzero sum of convex functions is convex. To prove that $f_{2}^{TM}$ is a concave function, we derive the second-order condition, namely the Hessian matrix, as follows:
\begin{small}
  \begin{equation}\label{eq:proof-3-2}
    \begin{aligned}
      \nabla ^2f_{2}^{TM}=\frac{1}{\left( 1-\lambda _n\tau _{b}^{TM} \right) ^3}\left[ \begin{matrix}
        \frac{2\left( 1-\lambda _n\tau _{b}^{TM} \right) ^3}{\lambda _{n}^{3}}+\left( \tau _{b}^{TM} \right) ^3&		\tau _{b}^{TM} & 0\\
        \tau _{b}^{TM}&		\lambda _n & 0\\
        0& 0& 0\\
      \end{matrix} \right].
  \end{aligned}
  \end{equation}
\end{small}
  It is clear that $\nabla ^2f_{2}^{TM}\preceq 0$ since the determinants obey $D_i\geq0$ for odd $i$ and $D_i\leq0$ for even $i$, and we have $D_1=\frac{\partial ^2f_{2}^{TM}}{\partial \left( \tau _{b}^{TM} \right) ^2}>0$, $D_2\geq 0$ and $D_3=\left| \nabla ^2f_{2}^{TM} \right|\leq0$. Therefore, it can be concluded that $f_{2}^{TM}$ is a concave function. By exploiting the second-order condition, we can prove that the inequality constraints (\ref{eq:tdma-1-ee2}), (\ref{OP:ET-1-1}\textrm{a})-(\ref{OP:ET-1-1}\textrm{c}) are convex based on the second-order condition, while (\ref{OP:ET-1-1}\textrm{d}) has to be rendered convex as $\left[ \lambda _n\left( \tau _{b}^{TM\,\,} \right) _{\left( k \right)}+\tau _{b}^{ST\,\,}\left( \lambda _{n}^{\,\,} \right) _{\left( k \right)} \right] <1+\left( \lambda _n\tau _{b}^{TM\,\,} \right) _{\left( k \right)}$, where the subscript $(k)$ denotes the value obtained from the $k$th iteration.
  
\end{Proof}
\end{appendices}
\ifCLASSOPTIONcaptionsoff
  \newpage
\fi

\bibliographystyle{IEEEtran}

\bibliography{ref}
\vspace{-5mm}
\begin{IEEEbiography}[{\includegraphics[width=1in,height=1.33in]{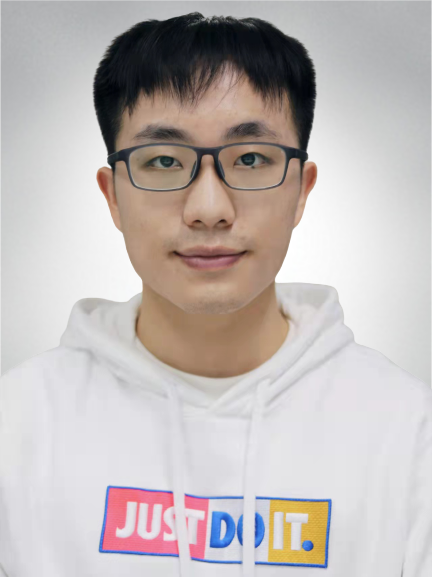}}]{\textbf{Zhengru Fang}} (S'20) received his B.S. degree in electronics and information engineering (with honors) from the Huazhong University of Science and Technology (HUST), Wuhan, China in 2019. He is currently pursuing the M.S. degree in electronics and communication engineering from Tsinghua University, Beijing, China. He has been serving as a Reviewer for IEEE JSAC, IEEE IoTJ, IEEE Systems Journal, IEEE Access, IEEE GLOBECOM and IEEE ICCC. His research interests lie in the area of Internet of Underwater Things, age of information, queueing theory and mobile edge computing.
  \vspace{-6mm}
  \end{IEEEbiography}
  \begin{IEEEbiography}[{\includegraphics[width=1in,height=1.33in]{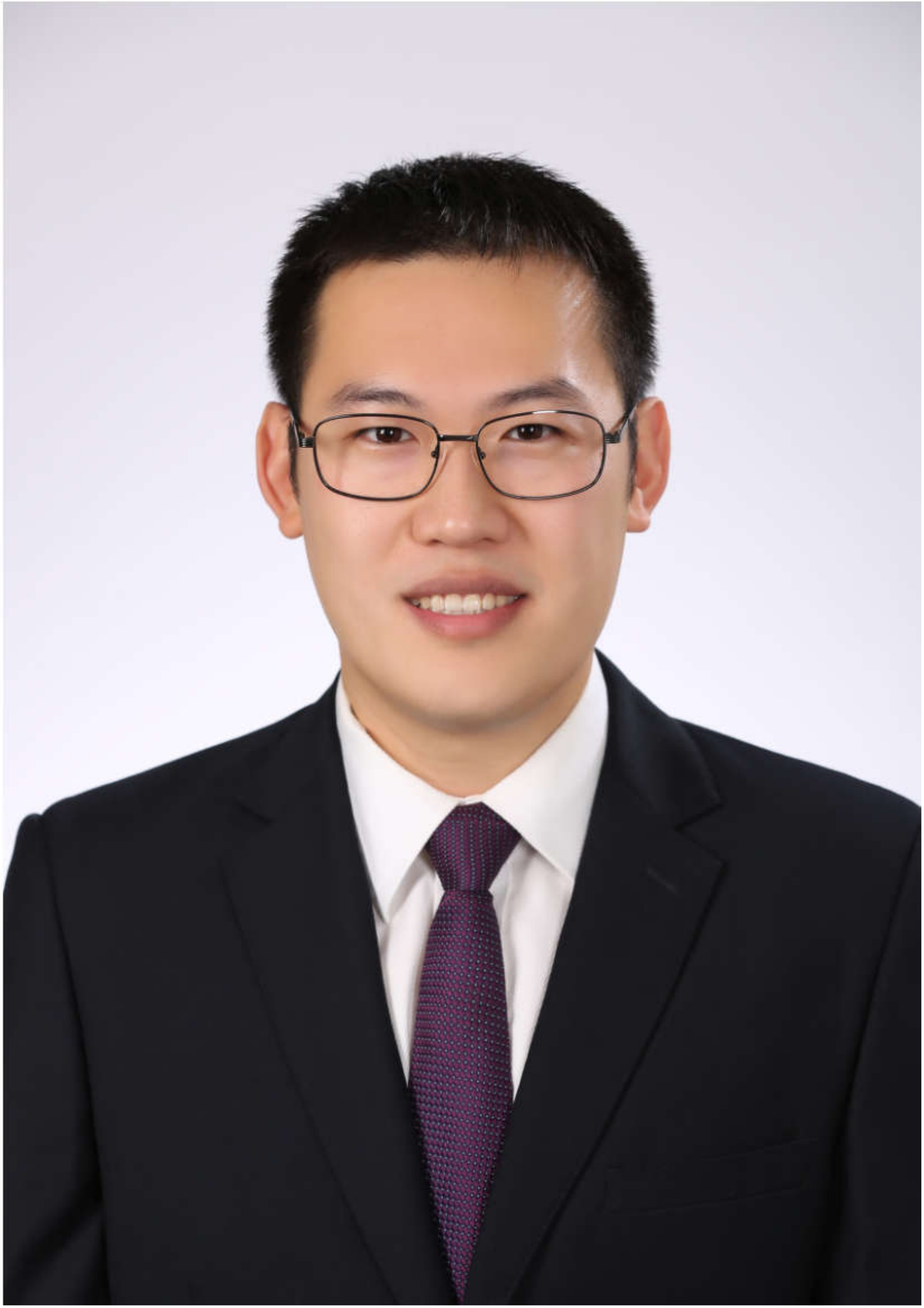}}]{\textbf{Jingjing Wang}} (S'14-M'19-SM’21) received his B.S. degree in Electronic Information Engineering from Dalian University of Technology, Liaoning, China in 2014 and the Ph.D. degree in Information and Communication Engineering from Tsinghua University, Beijing, China in 2019, both with the highest honors. From 2017 to 2018, he visited the Next Generation Wireless Group chaired by Prof. Lajos Hanzo, University of Southampton, UK. Dr. Wang is currently an associate professor at School of Cyber Science and Technology, Beihang University. His research interests include AI enhanced next-generation wireless networks, swarm intelligence and confrontation. He has published over 100 IEEE Journal/Conference papers. Dr. Wang was a recipient of the Best Journal Paper Award of IEEE ComSoc Technical Committee on Green Communications \& Computing in 2018, the Best Paper Award of IEEE ICC and IWCMC in 2019.
    \end{IEEEbiography}
    \begin{IEEEbiography}[{\includegraphics[width=1in,height=1.33in]{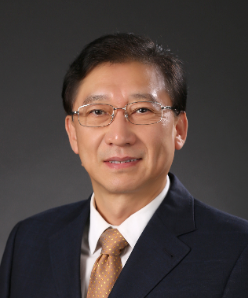}}]{\textbf{Yong Ren}} (M’11-SM’16) received his B.S, M.S and Ph.D. degrees in electronic engineering from Harbin Institute of Technology, China, in 1984, 1987, and 1994, respectively. He worked as a post doctor at Department of Electrical Engineering, Tsinghua University, China from 1995 to 1997. Now he is a full professor of Department of Electronic Engineering and serves as the director of the Complexity Engineered Systems Lab in Tsinghua University. He has authored or co-authored more than 400 technical papers in the area of computer network and mobile telecommunication networks. He has served as a reviewer of more than 40 international journals or conferences. His current research interests include complex system theory and its applications to the optimization of the Internet, Internet of Things and ubiquitous network, cognitive networks and cyber-physical systems.
    \end{IEEEbiography}
    \begin{IEEEbiography}[{\includegraphics[width=1in,height=1.33in]{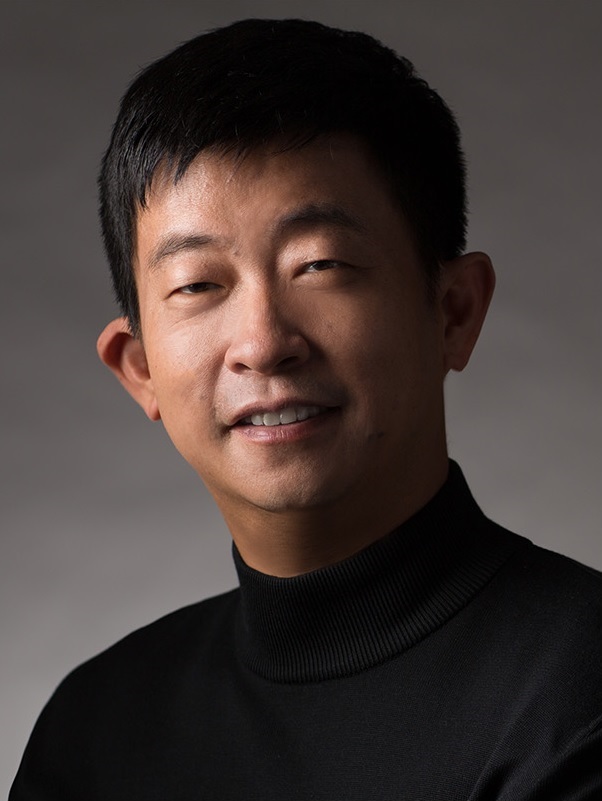}}]{\textbf{Zhu Han}} (S’01–M’04-SM’09-F’14) received the B.S. degree in electronic engineering from Tsinghua University, in 1997, and the M.S. and Ph.D. degrees in electrical and computer engineering from the University of Maryland, College Park, in 1999 and 2003, respectively. \par
      From 2000 to 2002, he was an R\&D Engineer of JDSU, Germantown, Maryland. From 2003 to 2006, he was a Research Associate at the University of Maryland. From 2006 to 2008, he was an assistant professor at Boise State University, Idaho. Currently, he is a John and Rebecca Moores Professor in the Electrical and Computer Engineering Department as well as in the Computer Science Department at the University of Houston, Texas. His research interests include wireless resource allocation and management, wireless communications and networking, game theory, big data analysis, security, and smart grid.  Dr. Han received an NSF Career Award in 2010, the Fred W. Ellersick Prize of the IEEE Communication Society in 2011, the EURASIP Best Paper Award for the Journal on Advances in Signal Processing in 2015, IEEE Leonard G. Abraham Prize in the field of Communications Systems (best paper award in IEEE JSAC) in 2016, and several best paper awards in IEEE conferences. Dr. Han was an IEEE Communications Society Distinguished Lecturer from 2015-2018, AAAS fellow since 2019 and ACM distinguished Member since 2019. Dr. Han is 1\% highly cited researcher since 2017 according to Web of Science. Dr. Han is also the winner of 2021 IEEE Kiyo Tomiyasu Award, for outstanding early to mid-career contributions to technologies holding the promise of innovative applications, with the following citation: ``for contributions to game theory and distributed management of autonomous communication networks.''
    \end{IEEEbiography}
    
   \begin{IEEEbiography}[{\includegraphics[width=1in,height=1.33in]{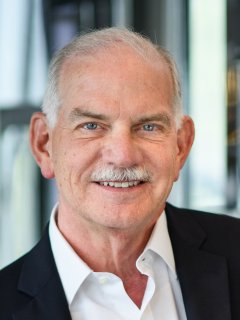}}]{\textbf{H. Vincent Poor}} (S’72, M’77, SM’82, F’87) received the Ph.D. degree in EECS from Princeton University in 1977.  From 1977 until 1990, he was on the faculty of the University of Illinois at Urbana-Champaign. Since 1990 he has been on the faculty at Princeton, where he is currently the Michael Henry Strater University Professor. During 2006 to 2016, he served as the dean of Princeton’s School of Engineering and Applied Science. He has also held visiting appointments at several other universities, including most recently at Berkeley and Cambridge. His research interests are in the areas of information theory, machine learning and network science, and their applications in wireless networks, energy systems and related fields. Among his publications in these areas is the forthcoming book \textit{Machine Learning and Wireless Communications}. (Cambridge University Press). Dr. Poor is a member of the National Academy of Engineering and the National Academy of Sciences and is a foreign member of the Chinese Academy of Sciences, the Royal Society, and other national and international academies. He received the IEEE Alexander Graham Bell Medal in 2017.
  \end{IEEEbiography}
  \begin{IEEEbiography}[{\includegraphics[width=1in,height=1.33in]{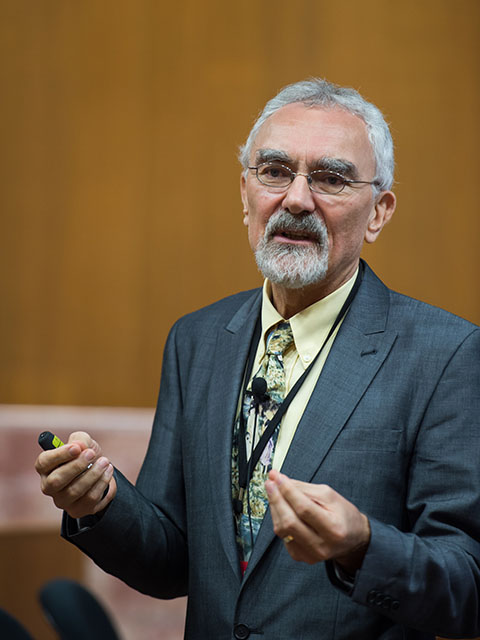}}]{\textbf{Lajos Hanzo}} (\url{http://www-mobile.ecs.soton.ac.uk},
    \url{https://en.wikipedia.org/wiki/Lajos_Hanzo}) (FIEEE'04) received his
    Master degree and Doctorate in 1976 and 1983, respectively from the
    Technical University (TU) of Budapest. He was also awarded the Doctor of
    Sciences (DSc) degree by the University of Southampton (2004) and
    Honorary Doctorates by the TU of Budapest (2009) and by the University
    of Edinburgh (2015).  He is a Foreign Member of the Hungarian Academy of
    Sciences and a former Editor-in-Chief of the IEEE Press.  He has served
    several terms as Governor of both IEEE ComSoc and of VTS.  He has
    published 2000+ contributions at IEEE Xplore, 19 Wiley-IEEE Press books
    and has helped the fast-track career of 123 PhD students. Over 40 of
    them are Professors at various stages of their careers in academia and
    many of them are leading scientists in the wireless industry. He is also
    a Fellow of the Royal Academy of Engineering (FREng), of the IET and of
    EURASIP. He is the recipient of the 2022 Eric Sumner Field Award.
  \end{IEEEbiography}  
\end{document}